\documentclass[12pt]{article}

\usepackage{colonequals}
\usepackage{epstopdf}
\usepackage{graphicx}
\usepackage{amssymb}
\usepackage{subfigure}
\usepackage{pstricks-add}
\usepackage{pst-slpe}
\usepackage{pst-coil}
\usepackage{color}
\usepackage{amsthm}
\usepackage{authblk}
\usepackage{placeins}
\usepackage[small]{caption}
\usepackage{amsmath}
\usepackage{wasysym}
\usepackage[bookmarks, colorlinks=true, linkcolor=blue, citecolor=green]{hyperref}

\title{Extremal Properties of Complex Networks}
\author[1,2]{Dionysios Barmpoutis}
\author[2]{Richard M. Murray}
\affil[1]{Computation and Neural Systems}
\affil[2]{Control and Dynamical Systems \bigskip}
\affil[ ]{California Institute of Technology\bigskip}
\affil[ ]{dionysios$@$caltech.edu}
\affil[ ]{murray$@$cds.caltech.edu}
\date{\today}                                        

\newtheorem{lemma}{Lemma}
\newtheorem{theorem}{Theorem}
\newtheorem{corollary}{Corollary}

\DeclareMathOperator*{\argmax}{arg\,max}

\begin{document}
\maketitle
\begin{abstract}
We describe the structure of connected graphs with the minimum and maximum average distance, radius, diameter, betweenness centrality, efficiency and resistance distance, given their order and size.
We find tight bounds on these graph qualities for any arbitrary number of nodes and edges and analytically derive the form and properties of such networks.
\end{abstract}
\section{Introduction}

Many complex systems can be described as interconnections of simpler elements, which in turn can be analyzed abstractly as graphs.
We are interested in the structural properties of these networks, regardless of the nature of their individual parts.
This allows the results developed in this paper to be applicable in a wide range of different disciplines, such as neuroscience, biology, social sciences and engineering. 
The properties that are of general interest are the average distance,  betweenness centrality, radius, diameter, efficiency and the graph resistance.
Depending on the application we usually want to minimize or maximize one or more of the above, because they are directly implicated in some performance metric of the network.
They are correlated with how fast the system responds to different input stimuli \cite{Strogatz2001} and how robust it is to the failure of individual subsystems, due to random failures or targeted attacks \cite{LoadDistribution},\cite{ErrorAttackTolerance},\cite{BreakdownIntentionalAttack}.
They also indicate how efficient message propagation is across a network \cite{EfficiencyDefinitionPaper}, how easy it is for dynamical processes that require global coordination and information flow (like synchronization or computation) to take place, and how reliable a transmitted message is in the presence of noise \cite{Strogatz2001}.
Although these structural properties do not take into account the specifics of the various systems, focusing on the structural patterns of the network architecture can give a valuable insight on how to optimize the network function, while obeying other constraints.

\section{Preliminaries}

This section provides a brief introduction to the notions from graph theory that are used throughout this study.
A \textit{graph} (also called a \textit{network}) is an ordered pair $\mathcal{G}=(\mathcal{V},\mathcal{E})$ comprised of a set $\mathcal{V}=\mathcal{V}(\mathcal{G})$ of \textit{vertices} together with a set $\mathcal{E}=\mathcal{E}(\mathcal{G})$ of \textit{edges} that are unordered 2-element subsets of $\mathcal{V}$.
Two vertices $u$ and $v$ are called \textit{neighbors} if they are connected through an edge ($(u,v)\in \mathcal{E}$) and we write $u-v$, otherwise we write $u \notslash v$.
All graphs in this article are \textit{simple}, meaning that all edges connect two different vertices, there is at most one edge connecting any pair of vertices, and edges have no direction.
The \textit{neighborhood} $\mathcal{N}_{u}$ of a vertex $u$ is the set of its neighbors.
The \textit{degree}  of a vertex is the number of its neighbors.
A vertex is said to have \textit{full degree} if it is connected to every other vertex in the network.
A network is \textit{assortative} with respect to its degree distribution when the vertices with large degrees are connected to others that have large degrees.
When vertices with small degrees connect to vertices with large degrees and vice versa, then the network is called \textit{disassortative}.
The \textit{order} $N$ of a graph is the number of its vertices, $N=|\mathcal{V}|$.
A graph's \textit{size} (denoted by $m=|\mathcal{E}|$), is the number of its edges.
We will denote a graph $\mathcal{G}$ of order $N$ and size $m$ as $\mathcal{G}(N,m)$ or simply $\mathcal{G}_{N,m}$.
A \textit{complete graph} is a graph in which each vertex is connected to every other.
The \textit{edge density} of a graph is defined as $\rho=m/{N \choose 2}$, representing the number of present edges, as a fraction of largest possible number of edges, which is the size of a complete graph.
A \textit{clique} in a graph is a subset of its vertices such that every vertex pair in the subset is connected.
The \textit{clique order} is the number of vertices that belong to it.
A \textit{path} is a sequence of consecutive edges in a graph and the length of the path is the number of edges traversed.
The \textit{distance} between two vertices $u$ and $v$, usually denoted by $d=d(u,v)$, is the length of the shortest path that connects these two vertices.
A \textit{cycle} is a closed (simple) path, with no other repeated vertices or edges other than the starting and ending nodes.
A \textit{full cycle} is a cycle that includes all the vertices of the network.
A graph is \textit{connected} if for every pair of vertices $u$ and $v$, there is a path from $u$ to $v$.
Otherwise the graph is called \textit{disconnected}.
We will be focusing exclusively on connected graphs, given that every disconnected graph can be analyzed as the sum of its connected \textit{components}.
If the distance between $u$ and $v$ is equal to $k$, then these vertices are called \textit{$k-$neighbors}, and the set of all pairs in the graph that are $k-$neighbors is denoted by $\mathcal{E}_{k}$.
The \textit{eccentricity} of a vertex $u$ is the maximum distance of $u$ from any other vertex in the graph.
A \textit{central vertex} of a graph is a vertex that has eccentricity smaller or equal to any other node.
A network may have many central vertices, all of which are considered its centers.
The eccentricity of a central vertex is called the graph \textit{radius}.
The graph \textit{diameter} is defined as the maximum of the distances among all vertex pairs in the network.
A \textit{tree} is a graph in which any two vertices are connected by exactly one path.
A \textit{cut} is a partition of the vertices of a graph into two disjoint subsets.
A \textit{cut set} of the cut is the set of edges whose end points are in different subsets of the cut.
A \textit{cut vertex} of a connected graph is a vertex that if removed, (along with all edges incident with it) produces a graph that is disconnected.
An edge is \textit{rewired} when we change the vertices it is adjacent to.
A \textit{single rewiring} takes place when we change one of the vertices that is adjacent to it, and a \textit{double rewiring} when we change both of them.
A subgraph $\mathcal{H}$ of a graph $\mathcal{G}$ is called \textit{induced} if $\mathcal{V}(\mathcal{H}) \subseteq \mathcal{V}(\mathcal{G})$ and for any pair of vertices $u$ and $v$ in $\mathcal{V}(\mathcal{H})$, $(u,v)\in \mathcal{E}(\mathcal{H})$ if and only if $(u,v) \in \mathcal{E}(\mathcal{G})$.
 In other words, $\mathcal{H}$  is an induced subgraph of $\mathcal{G}$ if it has the same edges that appear in $\mathcal{G}$ over the same vertex set. 
 Finally, two graphs $\mathcal{G}$ and $\mathcal{H}$ are called isomorphic if there exists a bijective function $f:\mathcal{V(\mathcal{G})} \to \mathcal{V(\mathcal{H})}$ such that 
 \begin{equation}
 (u,v) \in \mathcal{E(\mathcal{G})} \iff (f(u),f(v)) \in \mathcal{E(\mathcal{H})}.
 \end{equation}
 Two graphs that are isomorphic have by definition the same order and size, and are considered identical.
 
\section{Networks with the Minimum and Maximum Average Shortest Path Length}

\subsection{Minimum Average Distance}

The average distance of a network is an important property, since it is a direct indicator of how different parts of the network communicate, and exchange information.
A small average distance is a proxy for improved synchronizability, efficient computation and signal propagation across the network \cite{Strogatz2001}.
In this section, we will analytically compute the minimum average distance of a graph of fixed order and size, and find sufficient conditions in order to achieve that minimum.

\begin{lemma}
The average distance of a graph is a strictly decreasing function of its size.
If we start with graph $\mathcal{G}=\mathcal{G}(N,m)$ with average distance $\bar{D}(\mathcal{G}) $, and introduce one additional edge, the new graph $\mathcal{G'}=\mathcal{G'}(N,m+1)$ will have an average distance $\bar{D}(\mathcal{G'}) <\bar{D}(\mathcal{G}) $, for $N-1\leq m \leq {N\choose 2}-1$.
\label{AverageDistanceDecreasingFunction}
\end{lemma}
\begin{proof}
The additional edge will connect two previously non-neighboring vertices $s$ and $t$, changing their distance to $d'(s,t)=1$.
Since they were not connected before, their distance was $d(s,t)\geq 2$, so $d'(s,t)<d(s,t).$
For every other pair of vertices $u$ and $v$, the new edge can only create new shortest paths, so $d'(u,v) \leq d(u,v)$.
The total average shortest path length of the new graph is:
\begin{equation}
\bar{D}(\mathcal{G'})=\frac{1}{{N \choose 2}}\sum _{\substack {(u,v) \in \mathcal{V}^{2}(\mathcal{G'}) \\ u \neq v}} d'(u,v) < \frac{1}{{N \choose 2}}\sum _{\substack {(u,v) \in \mathcal{V}^{2}(\mathcal{G}) \\ u \neq v}} d(u,v) =\bar{D}(\mathcal{G}) .
\end{equation}
\end{proof}

\begin{lemma}
The star graph is the only tree of order $N$ that has the smallest average distance equal to $\bar{D}_{star}=2-\frac{2}{N} $.
\end{lemma}
\begin{proof}
A tree has exactly $N-1$ edges among its $N$ vertices.
There will be exactly $N-1$ pairs of vertices with distance $d=1$, and  ${N-1 \choose 2}$ vertex pairs that are not connected, with distances $d(u,v) \geq 2$.
The star graph achieves this lower bound, and has the minimum possible average distance.
\begin{equation}
\bar{D}_{star}=\frac{1}{{N \choose 2}}\sum _{\substack {(u,v) \in \mathcal{V}^{2} \\ u \neq v}} d'(u,v) = \frac{1}{{N \choose 2}} \left( N-1+2 {N-1 \choose 2}\right)=2-\frac{2}{N}.
\end{equation}
It is also unique: If a tree is not a star, there is no vertex that is connected to all the remaining vertices.
In this case, there are at least two vertices with distance $d \geq 3$, since in every tree there is a unique path connecting each vertex pair, and at the same time the number of neighboring vertices is the same as in the star graph.
\end{proof}

Using the same method as above, we can find the smallest average distance of a graph with $N$ vertices and $m$ edges, which we denote as $\bar{D}_{min}(N,m)$.

\begin{theorem}
The minimum possible average distance of a graph $\mathcal{G}(N,m)$  is equal to $\bar{D}_{min}(N,m)=2-\frac{m}{{N \choose 2}} $.
\end{theorem}
\begin{proof}
The graph $\mathcal{G}(N,m)$  has $m$ pairs of vertices with distance exactly $1$, and the rest of the pairs of vertices $(u,v)$ have distances $d(u,v) \geq 2$.
Consequently, its average distance is
\begin{equation}
\mathcal{L}_{\mathcal{G}} \geq \frac{m+ 2\left({N\choose 2} -m \right)}{{N \choose 2}}= 2-\frac{m}{{N \choose 2}}.
\label{SmallestDistance}
\end{equation}
This lower bound can always be achieved.
A connected graph $\mathcal{G}(N,m)$ with at least one vertex with degree $d=N-1$ has the star graph as an induced subgraph, so all non-neighboring vertices will have distance equal to $2$. 
All connected vertices have distance equal to $1$, leading to the lower bound of equation \eqref{SmallestDistance}.
\end{proof}

\begin{corollary}
If a graph $\mathcal{G}$ has at least one vertex pair $(u,v)$ with distance $d(u,v) \geq 3$, then its average distance is $\mathcal{L}_{\mathcal{G}} >\bar{D}_{min}(N,m)$.
\label{NecessaryConditionForNonneighbors}
\end{corollary}
\begin{proof}
The number of pairs with distance $1$ is fixed, equal to the graph's size. 
All other vertices have a distance of at least $2$, and the minimum is achieved when \textit{all} non-neighboring pairs have distance equal to $2$.
\end{proof}

The next three corollaries present sufficient conditions for a graph to have the smallest average shortest path length.

\begin{corollary}
In a network with the smallest average distance, all vertex pairs are either connected, or connected to a common third vertex.
\label{MinDistanceNecessaryCondition1}
\end{corollary}

\begin{corollary}
A cut of a minimum average distance graph $\mathcal{G}$ divides its vertices into two disjoint sets where, in at least one of the sets, all vertices have at least one neighbor in the other.
\label{MinDistanceNecessaryCondition2}
\end{corollary}
\begin{proof}
Assume that in both sets of a graph $\mathcal{G}$ there is at least one vertex which has no neighbors to the other set.
The distance between these two vertices is at least $3$, and according to Corollary  \ref{NecessaryConditionForNonneighbors},  graph $\mathcal{G}$ will not the smallest possible average distance.
\end{proof}

\begin{corollary}
Assume that graph $\mathcal{G}$ of order $N$ has the smallest average distance. 
The average degree $\bar{g}_{_{\mathcal{N}_{u}}}$ of the neighbors of vertex $u$ with degree $d_{u}$ satisfies the inequality
\begin{equation}
\bar{g}_{_{\mathcal{N}_{u}}}\geq \frac{N-1}{d_{u}}.
\end{equation}
\label{MinDistanceNecessaryCondition3}
\end{corollary}
\begin{proof}
Since every vertex $u$ of $\mathcal{G}$ has distance exactly $2$ with all its non-neighbors, the vertices in its neighbor set $\mathcal{N}_{u}=\{V_{1}, \dots ,V_{d_{u}}\}$ should be connected to all the remaining vertices.
In other words, all the remaining $N-1-d_{u}$ vertices of the graph should have at least one common neighbor with $u$.
Each neighbor $V_{k}$ of $u$ with degree $g_{k}$ has $g_{k}-1$ neighbors other than $u$, some of which may belong to $\mathcal{N}_{u}$.
If we add up the neighbors of all these vertices excluding $u$, we get:
\begin{equation}
\begin{aligned}
\sum _{k \in \mathcal{N}_{u}} (g_{k}-1) &\geq N-1-d_{u} \\
\sum _{k \in \mathcal{N}_{u}} g_{k} &\geq N-1 \\
d_{u} \bar{g}_{_{\mathcal{N}_{u}}}&\geq N-1 \\
\bar{g}_{_{\mathcal{N}_{u}}}&\geq \frac{N-1}{d_{u}}.
\end{aligned}
\end{equation}
\end{proof}

\begin{corollary}
Networks that have the smallest  possible average shortest path length are disassortative with respect to their degrees.
\end{corollary}
\subsection{Maximum Average Shortest Path Length}

The networks with the largest average distance have a very different architecture.
They consist of two distinct connected subgraphs, and if we remove any edge, the network either becomes disconnected, or the previously connected vertices become second neighbors. 

\begin{lemma}
Assume that a vertex $u$ with degree $d_{u}$ is added to a network, with its neighbor set being $\mathcal{N}_{u}$.
Rewiring edges of $\mathcal{G}$ such that they connect previously non-neighboring vertices in $\mathcal{N}_{u}$ cannot decrease its eccentricity or the average distance of $u$ with the other vertices in the network.
\label{CliqueMaxDistanceLemma}
\end{lemma}
\begin{proof}
Connecting any two vertices in $\mathcal{N}_{u}$ will not change the distance of $u$ with any of them.
Furthermore, disconnecting a pair of vertices, at least one of which is not in $\mathcal{N}_{u}$ can only increase the distance of $u$ with any of the vertices that do not belong to $\mathcal{N}_{u}$.
\end{proof}
More generally, connecting two non-neighboring vertices has the smallest impact on their average distance if they have a common neighbor.
Rewiring an edge in $\mathcal{G}$ will increase the distance of the initially connected pair $(u_{1},v_{1})$ to $d$, and decrease the distance of the new pair of vertices $(u_{2},v_{2})$ with a common neighbor by $1$.
The overall difference will be
\begin{equation}
\begin{aligned}
\Delta d(u_{1},v_{1})+\Delta d(u_{2},v_{2}) &=(d_{new}(u_{1},v_{1})-d_{old}(u_{1},v_{1})) \\
& \qquad \qquad +(d_{new}(u_{2},v_{2})-d_{old}(u_{2},v_{2}))\\
&=d-2 \\
&\geq 0.
\end{aligned}
\label{RewiringToSecondNeighbors}
\end{equation}

Combining Lemma \ref{CliqueMaxDistanceLemma} with equation \eqref{RewiringToSecondNeighbors}, we can easily see that for a fixed neighborhood $\mathcal{N}_{u}$ of a vertex $u$, we can increase the eccentricity of $u$ and at the same time the average distance of the graph it belongs to, simply by rewiring edges to connect vertices in $\mathcal{N}_{u}$, until they form a clique.

\begin{lemma}
All connected graphs of order $N \geq 2$ and size ${N-1 \choose 2}+1 \leq m \leq {N \choose 2}$ have the same average distance, equal to 
\begin{equation}
\bar{D}(N,m)=2-\frac{m}{{N \choose 2}}.
\label{AlmostCompleteGraphsOnlyDistance}
\end{equation}
\label{AlmostCompleteGraphsDistanceLemma}
\end{lemma}
\begin{proof}
Assume that the largest clique in $\mathcal{G}$ consists of $C$ vertices, that we will call \textit{central} vertices.
The rest of the nodes belong to the set $\mathcal{P}$ of \textit{peripheral} vertices, with $|\mathcal{P}|=P$ and they form connections to the central vertices or among themselves.
Since $m \geq {N-1 \choose 2}+1$, every vertex in the graph is either a central or a peripheral vertex, and as a result
\begin{equation}
C+P=N.
\end{equation}
The average distance of equation \eqref{AlmostCompleteGraphsOnlyDistance} is equal to the minimum possible distance of a graph as in equation \eqref{SmallestDistance}, and it is achieved if and only if all non-neighboring  vertices have distance equal to $2$.
The only way that the network will not have an average distance equal to $\bar{D}_{min}$ is when there is a pair of vertices $A$ and $B$ with shortest path length of at least $3$.
If there exist two such vertices, then from equation \eqref{SmallestDistance} and  Corollary \ref{NecessaryConditionForNonneighbors} we conclude that the maximum average distance of the graph will be
\begin{equation}
D(\mathcal{G})>2-\frac{m}{{N \choose 2}}.
\end{equation}
The central vertices are by definition fully connected to each other, and any peripheral vertex has distance two with all the central vertices it is not connected with.
So, the only case where two non-neighboring vertices do not have any common neighbors is when both of them are peripheral vertices.
We will now show that this is not possible.

For every peripheral vertex $u$, there are $\gamma _{u}$ central vertices that are \textit{not} connected to it.
Also, let $h$ be the total number of non-neighboring peripheral vertices.
The total number of non-neighboring vertex pairs is
\begin{equation}
\gamma = h + \sum _{u \in \mathcal{P}} \gamma _{u} 
\end{equation}
with 
\begin{equation}
\begin{aligned}
\gamma &= {N \choose 2}-m \\
& \leq {N \choose 2}-{N-1 \choose 2} -1 \\
&= N-2.
\end{aligned}
\end{equation}

In addition,
\begin{equation}
h\geq 1
\end{equation}
since $A$ and $B$ are not connected.
Combining all the equations above:
\begin{equation}
\begin{aligned}
h+\sum _{u \in \mathcal{P} } \gamma _{u} \leq N-2 &\implies \sum _{u \in \mathcal{P} } \gamma _{u} \leq N-3 \\
&\implies \gamma _{A} + \gamma _{B} +\sum _{ \substack{u \in \mathcal{P} \\ u \neq A, u \neq B }} \gamma _{u} \leq N-3.
\end{aligned}
\label{TwoPeripheralVerticesNonneighborInequality}
\end{equation}
Every peripheral vertex in $\mathcal{P}$ has at least one central vertex
that it is \textit{not} connected to, so
\begin{equation}
\gamma_{u} \geq 1 \quad \forall u \in \mathcal{P}
\end{equation}
and 
\begin{equation}
\sum _{ \substack{u \in \mathcal{P} \\ u \neq A, u \neq B }} \gamma _{u} \geq P-2.
\end{equation}
Based on the last two inequalities combined with inequality \eqref{TwoPeripheralVerticesNonneighborInequality}, we can derive an upper bound for the sum of $\gamma_{A}$ and $\gamma _{B}$:
\begin{equation}
\begin{aligned}
\gamma _{A} + \gamma _{B} &\leq N-P-1 \\
&\leq N-3
\end{aligned}
\end{equation}
because $P\geq 2$.
But $A$ and $B$ have by assumption no common neighbors in the clique or among any peripheral vertices, which means that
\begin{equation}
\gamma _{A} + \gamma _{B} \geq N-2
\end{equation}
which is clearly a contradiction.
\end{proof}

\begin{figure}[tb]
\subfigure[]{
\psscalebox{0.3}{
\begin{pspicture}(-5,-7)(16,8)
{
\cnodeput[fillstyle=solid,fillcolor=blue](3.12,3.91){A}{\strut}
\cnodeput[fillstyle=solid,fillcolor=blue](-1.11,4.87){B}{\strut}
\cnodeput[fillstyle=solid,fillcolor=blue](-4.5,2.17){C}{\strut}
\cnodeput[fillstyle=solid,fillcolor=blue](-4.5,-2.17){D}{\strut}
\cnodeput[fillstyle=solid,fillcolor=blue](-1.11,-4.87){E}{\strut}
\cnodeput[fillstyle=solid,fillcolor=blue](3.12,-3.91){F}{\strut}
\cnodeput[fillstyle=solid,fillcolor=green](5,0){G}{\strut}
\cnodeput[fillstyle=solid,fillcolor=green](7,0){H}{\strut}
\cnodeput[fillstyle=solid,fillcolor=green](9,0){I}{\strut}
\cnodeput[fillstyle=solid,fillcolor=green](11,0){J}{\strut}
\cnodeput[fillstyle=solid,fillcolor=green](13,0){K}{\strut}
\cnodeput[fillstyle=solid,fillcolor=green](15,0){L}{\strut}
}
\ncline{-}{A}{B}
\ncline{-}{A}{C}
\ncline{-}{A}{D}
\ncline{-}{A}{E}
\ncline{-}{A}{F}
\ncline{-}{B}{C}
\ncline{-}{B}{D}
\ncline{-}{C}{D}
\ncline{-}{B}{E}
\ncline{-}{C}{E}
\ncline{-}{D}{E}
\ncline{-}{B}{F}
\ncline{-}{C}{F}
\ncline{-}{D}{F}
\ncline{-}{E}{F}
\ncline{-}{C}{G}
\ncline{-}{D}{G}
\ncline{-}{E}{G}
\ncline{-}{F}{G}
\ncline{-}{G}{H}
\ncline{-}{H}{I}
\ncline{-}{I}{J}
\ncline{-}{J}{K}
\ncline{-}{K}{L}
\end{pspicture}
}
\label{MaxDistanceGeneralCase}
}
\subfigure[]{
\psscalebox{0.4}{
\begin{pspicture}(-1,-4)(13,5)
{
\cnodeput(0,0){A}{\strut}
\cnodeput(1.5,0){B}{\strut}
\cnodeput(3,0){C}{\strut}
\cnodeput(4.5,0){D}{\strut}
\cnodeput(6,0){E}{\strut}
\cnodeput(7.5,0){F}{\strut}
\cnodeput(9,0){G}{\strut}
\cnodeput(10.5,0){H}{\strut}
\cnodeput(12,0){I}{\strut}
\cnodeput(13.5,0){J}{\strut}
\cnodeput(2,3){K}{\strut \boldmath$A$}
\cnodeput(9,3){L}{\strut \boldmath$B$}
\rput[bl](5,4.5){\rnode{EV}{\Large{Peripheral Vertices}}}
}
\ncline[nodesep=5pt]{->}{EV}{K}
\ncline[nodesep=5pt]{->}{EV}{L}
\ncline{-}{A}{B}
\ncline{-}{B}{C}
\ncline{-}{C}{D}
\ncline{-}{D}{E}
\ncline{-}{E}{F}
\ncline{-}{F}{G}
\ncline{-}{G}{H}
\ncline{-}{H}{I}
\ncline{-}{I}{J}
\ncloop[angleA=180,loopsize=2,arm=.5,linearc=.2]{-}{A}{J}
\ncline{-}{A}{K}
\ncline{-}{B}{K}
\ncline{-}{C}{K}
\ncline{-}{D}{K}
\ncline{-}{E}{L}
\ncline{-}{F}{L}
\ncline{-}{G}{L}
\ncline{-}{H}{L}
\ncline{-}{I}{L}
\ncline{-}{J}{L}
\psbrace[braceWidth=0.2pt,rot=90,nodesepB=0.5](-0.5,-0.5)(5,-0.5){\Large{$d_{A}$}}
\psbrace[braceWidth=0.1pt,rot=90,nodesepB=0.5](5.5,-0.5)(13.8,-0.5){\Large{$d_{B}$}}
\psbrace[nodesepA=-20mm,nodesepB=0.5,braceWidth=0.2pt,rot=90](-0.7,-2.3)(14.4,-2.3){\Large{(N-2)-Complete Graph}}
\end{pspicture}
}
\label{AlmostAlmostCompleteGraphFigure}
}
\caption{\textbf{(a)} The graph of order $N=12$ and size $m=24$, with the largest average shortest path length. It consists of a complete graph of order $C=6$ (blue), and a path graph of order $P=N-C=6$ (green). Four edges ($\alpha=4$) connect the complete subgraph to one of the two ends of the path graph. \textbf{(b)} A network with size $m={N-1 \choose 2}$ and largest possible average distance. Vertices $A$ and $B$ are the only vertices without any common neighbors, and $d_{A}+d_{B}=N-2$, the number of central vertices. }
\end{figure}
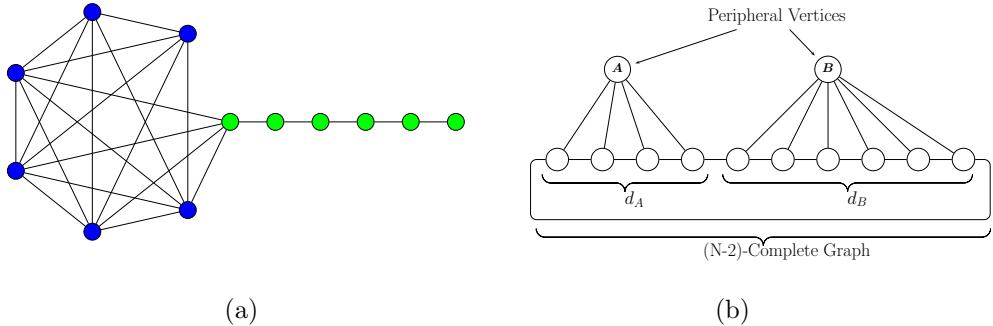

\begin{corollary}
There are exactly $\Big \lfloor  \frac{N-2}{2} \Big \rfloor$ non-isomorphic graphs of order $N$ and size $m={N-1 \choose 2}$ with the largest possible average distance, equal to 
\begin{equation}
\bar{D}_{max}(N,m)=2-\frac{m-1}{{N \choose 2}}.
\end{equation}
All other graphs of the same order and size have the minimum possible average distance among their vertices, equal to 
\begin{equation}
\bar{D}_{min}(N,m)=2-\frac{m}{{N \choose 2}}.
\end{equation} 
\end{corollary}
\begin{proof}
In a graph of size $m={N-1 \choose 2}$, the total number of missing edges among all the pairs of vertices is
\begin{equation}
\gamma= {N \choose 2}-{N-1 \choose 2}=N-1.
\end{equation}
Keeping the same notation as before, we add up all the missing edges among the peripheral vertices, and among peripheral and central vertices.
\begin{equation}
h+\gamma _{A} + \gamma _{B} +\sum _{ \substack{u \in \mathcal{P} \\ u \neq A, u \neq B }} \gamma _{u}= N-1
\label{AlmostAlmostCompleteGraphEquation}
\end{equation}
under the constraints
\begin{equation}
\gamma _{A}+\gamma_{B} \geq N-P, \quad \sum _{ \substack{u \in \mathcal{P} \\ u \neq A,B }} \gamma _{u} \geq P-2 \quad \textrm{and} \quad h \geq 1.
\end{equation}
These inequalities can only be satisfied in equation \eqref{AlmostAlmostCompleteGraphEquation} if all variables are equal to their respective lower bounds, namely
\begin{equation}
\gamma _{A}+\gamma_{B}=N-P, \quad \sum _{ \substack{u \in \mathcal{P} \\ u \neq A,B }} \gamma _{u}=P-2 \quad \textrm{and} \quad h=1.
\end{equation}
The only unknown variable above is $P$.
Since $A$ and $B$ are not neighbors,  and there is only one ($h=1$) edge missing among peripheral vertices.
If we assume that $P\geq 3$, then $A$ and $B$ have $P-2$ common neighbors, all peripheral vertices that are connected to both of them.
This clearly contradicts our assumption.
So $A$ and $B$ are the only peripheral vertices and $P=2$.
Such a graph is shown in Figure \ref{AlmostAlmostCompleteGraphFigure}.
It is clear from the previous analysis that
\begin{equation}
d_{A}+d_{B}+\gamma _{A} + \gamma _{B}=2(N-2) \implies d_{A}+d_{B}=N-2
\end{equation}
with $d_{A},d_{B} \geq 1$ because the graph is connected.
Setting $d_{A}\leq d_{B}$ in order to count only non-isomorphic graphs, it is clear that there are exactly $\big \lfloor  \frac{N-2}{2} \big \rfloor$ pairs $d_{A},d_{B}$ that satisfy the last equation.
\end{proof}

\begin{theorem}
The graph of order $N$ and size $N-1 \leq m \leq {N-1 \choose 2}$ with the largest average distance among its vertices consists of a complete subgraph of order $C$, and a path subgraph of order $P=N-C$.
The two subgraphs are connected through $\alpha$ edges, as shown in Figure \ref{MaxDistanceGeneralCase}.
In addition, the graph with the maximum average shortest path length is unique for $N-1 \leq m \leq {N-1 \choose 2}-1$.
\label{MaxDistanceMainTheorem}
\end{theorem}
\begin{proof}
Every arbitrary cut $S$ will produce two disjoint subgraphs, both of which need to be maximum distance graphs for the respective orders and sizes.
More formally, if $\mathcal{A}$ is the set of all networks  of all orders and sizes with the maximum possible average shortest path length and $\mathcal{H}$ is an induced subgraph of a graph $\mathcal{G}$, then
\begin{equation}
\mathcal{G} \in \mathcal{A} \iff \mathcal{G-H} \in \mathcal{A}  \qquad \forall \quad \mathcal{H} \subseteq \mathcal{G}.
\label{EverySubsetLargestAverageDistance}
\end{equation}
The above equation is a necessary and sufficient condition for maximum average distance.
If it does not hold for some subgraph $\mathcal{J}\subseteq \mathcal{G}$, then we would be able to rearrange the edges in it, so that the average distance among the vertices in the subgraph is increased.
Since this would also increase the average distance of $\mathcal{G}-\mathcal{J}$ with the vertices of $\mathcal{J}$, the overall average distance of $\mathcal{G}$ would increase.

Now suppose that we want to find the maximum average distance graph of order $N$.
According to the equation above, and setting one of the vertices $u$ as the chosen subgraph (of unit order), a graph with order $N$ and size $m$ has the largest possible average distance (in which case it is denoted $\mathcal{G}_{max}$) when
\begin{equation}
\begin{aligned}
\mathcal{G}_{max}(N,m) &= \argmax _{\mathcal{G} \in \mathcal{C}(N,m)}  \left[ \sum _{(u,v) \in \mathcal{V}^{2}(\mathcal{G})} d(u,v) \right], \\
\end{aligned}
\label{InductionEquation}
\end{equation}
where $\mathcal{C}(N,m)$ is the set of all possible connected graphs of order $N$ and size $m$.
But from equation \eqref{EverySubsetLargestAverageDistance}, and considering a subgraph $\mathcal{H}$ of order $1$, we can write the last condition as
\begin{equation}
\begin{aligned}
\mathcal{G}_{max}(N,m)=\max _{\mathcal{N}_{u}} \Bigg [ \mathcal{G}_{max}(N-1,m-|\mathcal{N}_{u}|) \cup \mathcal{H}(1,\mathcal{N}_{u}) \Bigg ].
\end{aligned}
\end{equation}

We will now find the neighborhood $\mathcal{N}_{u}$ of vertex $u$ in order to yield the graph with the largest average distance.
We will use induction.
For $N<4$, the theorem holds trivially.
For order $N=4$, it is easy to check that graphs of all sizes have the structure of the theorem.

Assume that all the maximum average distance graphs up to order $N_{0}$ and size $m_{0}$ have the same form described above, where 
\begin{equation}
N_{0}=N-1 \textrm{ and } N_{0}-1 \leq m_{0} \leq {N_{0} \choose 2}.
\end{equation}
It will be shown that all networks of order $N$ also have that same form, making use of equations \eqref{EverySubsetLargestAverageDistance} and \eqref{InductionEquation}.
If $d_{u}=1$, then we can connect it to the vertex $w$ with the largest average shortest path length.
In the resulting graph,  $u$ will now have the largest eccentricity and average distance to the other vertices.
At the same time the new graph will have the form stated in the theorem and its average distance to other vertices will be
\begin{equation}
\bar{D}_{u}=\sum _{\substack{v \in \mathcal{V}(\mathcal{G}) \\ v \neq u}} d(u,v)=1+\sum _{\substack{v \in \mathcal{V}(\mathcal{G}) \\ v \neq u,w}} (1+d(w,v)).
\end{equation}
If the degree of $u$ is equal to the order of the clique, the resulting graph will have the largest average distance if we connect it to all the vertices of the clique, as shown in Lemma \ref{CliqueMaxDistanceLemma}.
If $d_{u}$ is smaller than the order of the clique, then $u$ could be connected to clique vertices only, path vertices only, or a combination of both.
None of the above is an optimal configuration, since they do not satisfy condition \eqref{EverySubsetLargestAverageDistance}.
The same argument holds when $d_{u}$ is larger than the size of the clique.
In this case we can subtract the order of the clique $C$, and consider a new vertex with degree $d_{u}-C$, repeating the process if needed.
According to the above analysis, the new graph will either have the form stated in the theorem, or it will not have the largest average distance.

Finally for graphs with size  $N-1 \leq m \leq {N-1 \choose 2}$, the structure that yields the largest average distance is unique.
Using induction again, we see that for $N=4$, the claim holds.
For $N \geq 5$, the graph with maximum average distance is unique for $N-1$ by the induction hypothesis, and adding one extra vertex $u$ with $d_{u}=1$ or $d_{u}=C$ yields the same graph in both cases:
\begin{equation}
\mathcal{G}_{max}(N-1,m-C) \cup \mathcal{H}(1,C) \equiv \mathcal{G}_{max}(N-1,m-1) \cup \mathcal{H}(1,1).
\end{equation}
\end{proof}

Note that according to condition \eqref{EverySubsetLargestAverageDistance}, the network should have the same form no matter which subset of vertices we remove.
The form of a graph with the largest average distance as stated in Theorem  \ref{MaxDistanceMainTheorem} is one that satisfies that requirement.

The networks with the maximum average distance can be described as a combination of a \textit{type $I$ almost complete} subgraph \cite{MaxClusteringNetworks} and a path subgraph.
Since the only type of almost complete graph in this study is type $I$, we will refer to it simply as an almost complete graph, to avoid confusion.

We can now summarize the form of the networks with the largest average distance  for any number of edges.

\begin{corollary}
A graph $\mathcal{G}(N,m)$ with the largest average distance consists of a clique connected to a path graph as described in Theorem \ref{MaxDistanceMainTheorem} (see Figure \ref{MaxDistanceGeneralCase}) and is unique for $N-1 \leq m \leq {N-1 \choose 2}-1$.
If $m={N-1 \choose 2}$, then it consists of a complete subgraph of order $N-1$ and a vertex with degree $1$, or a clique of order $N-2$ and two peripheral vertices as shown in Figure \ref{AlmostAlmostCompleteGraphFigure}.
If $m \geq {N-1 \choose 2}+1$ then all graphs have the same average distance.
\end{corollary}

\begin{corollary}
Networks with the largest average shortest path length are assortative with regard to their degrees.
\end{corollary}

Note the difference between the networks with the smallest average distance and the largest average distance.
We can generally say that the average distance of a network is an increasing function of its assortativity.

The order of the clique and path subgraphs in a network with the largest average distance is computed below. 

\begin{corollary}
The average shortest path length among the vertices of a network with the largest possible average distance $\mathcal{G}_{max}(N,m)$ of order $N$ and size $m$, is equal to 
\begin{equation}
\bar{D}_{max}(N,m)=\frac{{C \choose 2}+{P+1 \choose 2}+(C-\alpha)P+{P+1\choose 3}}{{N \choose 2}},
\label{LargestDistance}
\end{equation}
where 
\begin{equation}
C= \Bigg \lfloor  \frac{3+\sqrt{9+8m-8N}}{2}  \Bigg \rfloor
\label{MaxDistanceGraphCliqueOrder}
\end{equation}
is the number of vertices that belong to the clique,
\begin{equation}
P= N-C
\label{MaxDistanceGraphPathOrder}
\end{equation}
is the number of vertices of the path subgraph and 
\begin{equation}
\alpha = m-P+1-{C \choose 2}
\label{CutEdgesBetweenPathAndClique}
\end{equation}
is the number of edges that connect the clique with the path graph.
\label{MaxDistanceCalculation}
\end{corollary}
\begin{proof}
We will find the lengths of the shortest paths among all vertices, add them, and finally divide them by their number to find the average.
First, we need to find the order of the clique.
Adding up all the edges of the network, we have
\begin{equation}
{C \choose 2} +\alpha +(P-1)=m.
\label{TotalNumberOfEdgesEquation}
\end{equation}
Replacing $P$ (total number of vertices is $C+P=N$),  we get
\begin{equation}
{C \choose 2} +\alpha +(N-C-1)=m
\end{equation}
where $C$ and $\alpha$ are integers satisfying the inequalities
\begin{equation}
1\leq C,P \leq N-1
\end{equation}
and 
\begin{equation}
1\leq \alpha \leq C-1
\end{equation}
respectively.
Solving for $C$:
\begin{equation}
C^{2}-3C+(2N-2m+2-2\alpha)=0.
\end{equation}
One way to find the solution of the second order equation above, is to set $\alpha$ equal to its smallest possible value, and solve for $C$, keeping in mind that it is always a positive integer.
As we add more edges, $\alpha$ is increasing with $C$ staying unchanged, until the vertex of the path subgraph is connected to all the vertices of the clique.
At this point, $C$ increases by one and $\alpha$ changes from $C-1$ to $1$.
We set $\alpha=1$, and taking into account that $C \in \mathbb{N}^{*}$,
\begin{equation}
C= \Bigg \lfloor  \frac{3+\sqrt{9+8m-8N}}{2}  \Bigg \rfloor.
\end{equation}
We can now compute the number of the vertices that do not belong to the clique, and the number of edges between the two subgraphs $\alpha$ from equation \eqref{TotalNumberOfEdgesEquation}.

The distance among each pair of the $C$ vertices of the clique is $1$, so the sum of the pairwise distances is 
\begin{equation}
D_{1}={C \choose 2}.
\end{equation}
The sum of the shortest path lengths of the path subgraph vertices to the clique vertices is 
\begin{equation}
\begin{aligned}
D_{2}&=\sum _{x=1}^{P} [ x\alpha +(x+1)(C-\alpha)  ]=\sum _{x=1}^{P} [ (C-\alpha) +xC  ] \\
&=P(C-\alpha)  +C {P+1\choose 2}.
\end{aligned}
\end{equation}
Finally, the sum of the shortest path lengths of nodes of the path subgraph is
\begin{equation}
D_{3}=\sum _{x=1}^{P} \sum _{y=1}^{x} (y-x) =\sum _{x=1}^{P} \sum _{z=0}^{x-1} z =\sum _{x=1}^{P} {x \choose 2} = {P+1 \choose 3}.
\end{equation}
Adding all the sums of all the shortest path lengths, and dividing by the total number of vertex pairs, we get 
\begin{equation}
\bar{D}_{max}(N,m)=\frac{{C \choose 2}+C{P+1 \choose 2}+P(C-\alpha)+{P+1\choose 3}}{{N \choose 2}}.
\end{equation}
\end{proof}

It is easy to show that when $m\geq {N-1 \choose 2}$, the formula for the minimum and maximum average distance give the same result for the average distance, in accordance with Lemma \ref{AlmostCompleteGraphsDistanceLemma}.
In that case, equation \eqref{LargestDistance} assumes that the network is an almost complete graph, but this graph has the same average distance as any other graph of the same order and size.

\begin{figure}[tb]
\begin{center}
\includegraphics[scale=0.35]{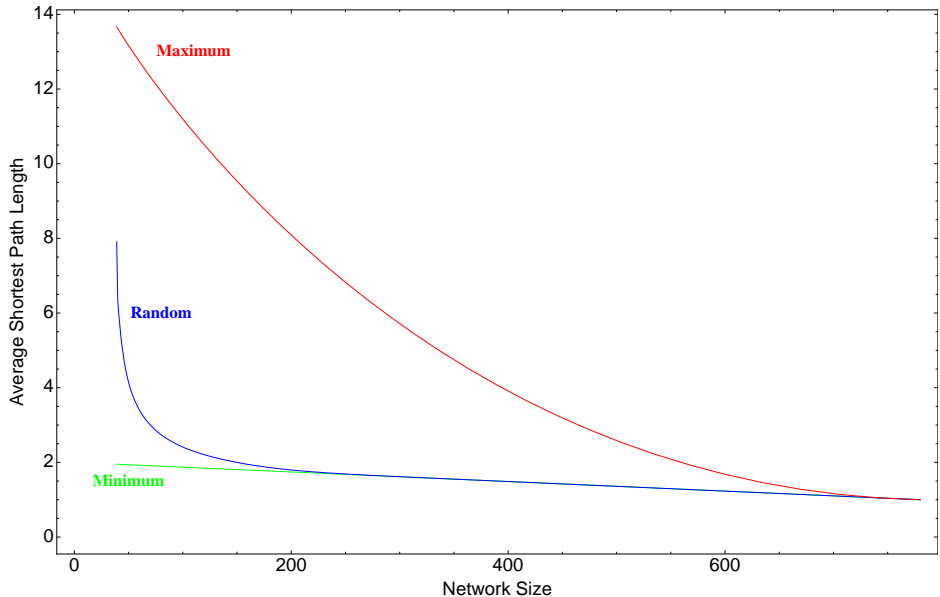}
\caption{Tight bounds on the average distance of a graph with $N=40$ vertices and $39\leq m \leq 780$ edges. These bounds have been computed analytically. The average shortest path length for random graphs has been estimated by finding the mean shortest path length of $10^{4}$ randomly generated graphs of the same order and size. The expected average distance of a random graph is very close to the minimum, even for relatively sparse networks.  For graphs with edge density $\rho>0.25$, it is virtually identical to the minimum one.}
\label{MinMaxDistanceBoundsExample}
\end{center}
\end{figure}

An example that shows the tight upper and lower bounds of the average distance of a graph with $N=40$ and $39\leq m \leq 780$ vertices is shown in Figure \ref{MinMaxDistanceBoundsExample}.

\FloatBarrier
\section{Betweenness Centrality}

The betweenness centrality of a vertex or an edge is a measure of how important this vertex or edge is for the communication among the different parts of the network.
It is based on counting the number of shortest paths among all pairs of vertices a given vertex or edge is a part of \cite{GraphTheoryBook}.
The vertex betweenness centrality is defined as
\begin{equation}
\mathcal{B}(u)=\sum _{\substack {(s,t) \in \mathcal{V}^{2}(\mathcal{G}) \\ s \neq u \neq t}} \frac{\sigma_{st}(u)}{\sigma_{st}},
\label{VertexBetweennessDefinition}
\end{equation}
where $\sigma _{st}$ is the number of shortest paths between vertices $s$ and $t$ and $\sigma _{st}(u)$ is the number of shortest paths between $s$ and $t$ that go through vertex $u$.
What equation \eqref{VertexBetweennessDefinition} computes is the total number of shortest paths of all the pairs of vertices in the graph that go through a given vertex $u$.
If there is more than one such path, we divide by their number, since they are assumed to be equally important.
The betweenness centrality of a vertex is sometimes normalized by the total number of all vertex pairs that we took into account for computing it, which is equal to ${N-1 \choose 2}$.
\begin{equation}
\mathcal{B}_{norm}(u)=\frac{1}{{N-1 \choose 2}}\sum _{\substack {(s,t) \in \mathcal{V}^{2}(\mathcal{G}) \\ s \neq u \neq t}} \frac{\sigma_{st}(u)}{\sigma_{st}}.
\end{equation}
The vertex betweenness is always nonnegative.
The only vertices with betweenness centrality equal to zero are the ones with degree equal to $1$.
In order to assess a network, we find the average for all vertices:
\begin{equation}
\mathcal{B}^{v}(\mathcal{G})=\frac{1}{N}\sum _{u \in \mathcal{V}(\mathcal{G})} \mathcal{B}(u).
\label{GraphVertexBetweennessCentralityDefinition}
\end{equation}
Networks with a large betweenness centrality usually have few vertices that play a major role in the communications among every other vertex.
Conversely, a small betweenness centrality indicates that all vertices are equally important or there are many different shortest paths among the various parts of the network.

The edge betweenness centrality is  similarly defined as the sum of the fraction of shortest paths of all vertex pairs in the network that go through a given edge:
\begin{equation}
\mathcal{B}(f)=\sum _{\substack {(s,t) \in \mathcal{V}^{2}(\mathcal{G}) \\ s\neq t}} \frac{\sigma_{st}(f)}{\sigma_{st}}
\label{EdgeBetweennessDefinition}
\end{equation}
where in this case $\sigma _{st}(f)$ is the number of shortest paths between $s$ and $t$ that go through edge $f$.
The edge betweenness centrality of the network is defined in the same manner as before:
\begin{equation}
\mathcal{B}^{e}(\mathcal{G})=\frac{1}{m}\sum _{f \in \mathcal{E}(\mathcal{G})} \mathcal{B}(f).
\end{equation}
The betweenness of an edge is always positive for a connected network.

The betweenness centrality of a graph is an important proxy of how robust the network is to random vertex or edge removals.
Removing a vertex or an edge with large betweenness centrality means that the communication among many vertex pairs will be affected, since they will now be forced to exchange information through alternative, possibly longer paths.
Graphs with large betweenness centralities are sensitive to random removal of a set of vertices or edges.
The vertex or edge betweenness centrality of a graph does not give any information about the centralities of different vertices or edges, which may have large variation among each other.
For networks with the same betweenness centrality, large variations among vertices or edges reveal a sensitivity to targeted attacks, since removing the most central vertices may significantly disrupt the network function.
In this section we show that the betweenness centrality of a graph is inherently related to its average shortest path length.

\begin{theorem}
The average betweenness centrality of a network $\mathcal{G}(N,m)$ is a linear function of its average distance, 
\begin{equation}
\mathcal{B}(\mathcal{G}) =\frac{(N-1)(\bar{D}(\mathcal{G})-1)}{2}.
\end{equation}
\label{BetweennessDistanceTheorem}
\end{theorem}
\begin{proof}
\begin{equation}
\begin{aligned}
\mathcal{B}(\mathcal{G}) &=\frac{1}{N}\sum _{u \in \mathcal{V}(\mathcal{G})} \mathcal{B}(u)
=\frac{1}{N}\sum _{u \in \mathcal{V}(\mathcal{G})} \sum_{\substack {(s,t) \in \mathcal{V}^{2}(\mathcal{G})\\ s\neq u\neq t}}\frac{\sigma_{st}(u)}{\sigma_{st}}\\
&=\frac{1}{2N}\sum _{u \in \mathcal{V}(\mathcal{G})} \sum_{\substack {s \in \mathcal{V}(\mathcal{G}) \\ s\neq u}} \sum_{\substack {t \in \mathcal{V}(\mathcal{G}) \\ t\neq u \\ t \neq s}} \frac{\sigma_{st}(u)}{\sigma_{st}}
=\frac{1}{2N} \sum _{\substack {s \in \mathcal{V}(\mathcal{G})}}  \sum _{\substack {t \in \mathcal{V}(\mathcal{G}) \\ t\neq s}}     \frac{1}{\sigma_{st}}  \sum _{\substack{ u \in \mathcal{V}(\mathcal{G}) \\ u\neq s\\ u \neq t}}  \sigma_{st}(u) \\
&=\frac{1}{2N} \sum _{s \in \mathcal{V}(\mathcal{G})}  \sum _{\substack {t \in \mathcal{V}(\mathcal{G}) \\ t\neq s}}     \frac{1}{\sigma_{st}} \sigma_{st}  \left(  |\mathcal{P}(s,t)|-1 \right)
=\frac{1}{2N} \sum _{s \in \mathcal{V}(\mathcal{G})}  \sum _{\substack {t \in \mathcal{V}(\mathcal{G}) \\ t\neq s}} \left( d(s,t)-1  \right) \\
&=\frac{1}{2N}   \left[ 2{N \choose 2} \bar{D}(\mathcal{G})  - 2 {N \choose 2} \right].
\end{aligned}
\end{equation}
Simplifying the last equation, the average betweenness centrality of a graph becomes the one stated in the theorem.
\end{proof}

It is worth mentioning that the average betweenness centrality of the network is only dependent on its size indirectly, through the average distance of the graph.
For a fixed order, the average betweenness centrality of a network decreases as we add new edges (see Lemma \ref{AverageDistanceDecreasingFunction}).

\begin{corollary}
A network has minimum (maximum) average betweenness centrality if and only if it has minimum (maximum) average distance.
The minimum possible average betweenness centrality of a graph of order $N$ and size $m$ is equal to 
\begin{equation}
\mathcal{B}_{min}(\mathcal{G})=\frac{N-1}{2} - \frac{m}{N}
\label{MinimumAverageBetweenness}
\end{equation}
and the maximum possible average betweenness centrality of such a graph is 
\begin{equation}
\mathcal{B}_{max}(\mathcal{G})=\frac{{C \choose 2}+C{P+1 \choose 2}+P(C-\alpha)+{P+1\choose 3}}{N}- \frac{N-1}{2}
\end{equation}
where $C,P$ and $\alpha$ are defined in equations \eqref{MaxDistanceGraphCliqueOrder}, \eqref{MaxDistanceGraphPathOrder} and \eqref{CutEdgesBetweenPathAndClique} respectively.
\label{MinMaxBetweennessCentralityCorollary}
\end{corollary}
\begin{proof}
The networks with the smallest or largest average betweenness centrality are exactly the graphs with the smallest or largest average distance respectively.
Replacing them from equations \eqref{SmallestDistance} and \eqref{LargestDistance},  the bounds for the average betweenness centrality of graphs follow.
\end{proof}

\begin{corollary}
The minimum sum of betweenness centralities of all the vertices of a network is equal to the number of vertices that are not neighbors.
\end{corollary}
\begin{proof}
From equations \eqref{GraphVertexBetweennessCentralityDefinition} and \eqref{MinimumAverageBetweenness}, we see that
\begin{equation}
\min _{\mathcal{G} \in \mathcal{C}(N,m)} \Bigg [ \sum _{u \in \mathcal{V}(\mathcal{G})} \mathcal{B}(u) \Bigg ]  =N \cdot \mathcal{B}_{min}(\mathcal{G})={N \choose 2}-m.
\end{equation}
\end{proof}

\begin{theorem}
The average edge betweenness centrality of a network is directly proportional to the average distance of the network, equal to
\begin{equation}
\mathcal{B}^{e}(\mathcal{G})=\frac{1}{m}{N \choose 2} \bar{D}(\mathcal{G}).
\end{equation}
The minimum  and maximum average edge betweenness centrality of a network of order $N$ and size $m$ are respectively
\begin{equation}
\mathcal{B}^{e}_{min}(N,m)=\frac{N(N-1)}{m}-1
\label{MinEdgeBetweenness}
\end{equation}
and
\begin{equation}
\mathcal{B}^{e}_{max}(N,m)=\frac{{C \choose 2}+C{P+1 \choose 2}+P(C-\alpha)+{P+1\choose 3}}{m}
\label{MaxEdgeBetweenness}
\end{equation}
where $C,P$ and $\alpha$ are the same as in equations \eqref{MaxDistanceGraphCliqueOrder}, \eqref{MaxDistanceGraphPathOrder} and \eqref{CutEdgesBetweenPathAndClique}.
\end{theorem}
\begin{proof}
We follow the same method as in the proof of the vertex betweenness centrality:
\begin{equation}
\begin{aligned}
\mathcal{B}_{e}(\mathcal{G}) &=\frac{1}{m}\sum _{e \in \mathcal{E}(\mathcal{G})} \mathcal{B}(e)
=\frac{1}{m}\sum _{e \in \mathcal{E}(\mathcal{G})} \sum_{\substack {(s,t) \in \mathcal{V}^{2}(\mathcal{G})\\ s\neq t}}\frac{\sigma_{st}(e)}{\sigma_{st}}\\
&=\frac{1}{m}\sum_ {(s,t) \in \mathcal{V}^{2}(\mathcal{G})} \sum _{e \in \mathcal{E}(\mathcal{G})} \frac{\sigma_{st}(e)}{\sigma_{st}}
=\frac{1}{m} \sum_ {(s,t) \in \mathcal{V}^{2}(\mathcal{G})} \frac{1}{\sigma_{st}}  \sum _{e \in \mathcal{E}(\mathcal{G})}  \sigma_{st}(e) \\
&=\frac{1}{m} \sum_ {(s,t) \in \mathcal{V}^{2}(\mathcal{G})} \frac{1}{\sigma_{st}} \sigma_{st}d(s,t) \\
&=\frac{1}{m}{N \choose 2} \bar{D}(\mathcal{G}).
\end{aligned}
\end{equation}
Replacing the average distance by its minimum and maximum bounds, we get equations \eqref{MinEdgeBetweenness} and \eqref{MaxEdgeBetweenness} respectively.
\end{proof}

\section{Efficiency}

The efficiency of a network (as defined in \cite{EfficiencyDefinitionPaper}) is a metric that shows how fast a signal travels on average in the network, assuming constant speed from one vertex to another.
It is the sum of the inverse distances of all vertex pairs in a network, normalized by the total number of such pairs:
\begin{equation}
\mathcal{F}(\mathcal{G})=\frac{1}{N(N-1)} \sum _{ \substack{u, v \in \mathcal{V}(\mathcal{G})\\ u \neq v} }  \frac{1}{d_{u,v}}.
\end{equation}

Network efficiency is also correlated with the fault tolerance of the network, in the sense of how the average distance of a network changes when one or more vertices are removed from the network.
It is has been used to assess the quality of neural, communication and transportation networks \cite{EfficiencyDefinitionPaper}.

Below we are going to show that the most and least efficient networks are the ones with the smallest and largest average distance among their individual parts.

\begin{theorem}
A graph $\mathcal{G}=\mathcal{G}(N,m)$ has the highest efficiency among all other graphs with the same order and size if and only if it is a graph of minimum average distance.
The highest efficiency of a network of $N$ vertices and $m$ edges is equal to 
\begin{equation}
\begin{aligned}
\mathcal{F}_{max}(N,m)=\frac{1}{2}+\frac{m}{N(N-1)}.
\end{aligned}
\label{MaxEfficiencyEquation}
\end{equation}

\end{theorem}
\begin{proof}
We assign a distance matrix to every graph, with its $(k,m)$ element being the distance between vertices $k$ and $m$.
For a graph $\mathcal{G}=\mathcal{G}_{N,m}$ with distance matrix $D$ and the minimum average distance, the sum of all the distances among all the pairs of vertices is smaller or equal to that of any other random graph $\mathcal{R}=\mathcal{R}_{N,m}$ with distance matrix $H$.
\begin{equation}
\sum _{k<m}d_{km} \leq  \sum _{k<m}h_{km}.
\end{equation}
The function to be maximized is convex, which means that the maximum lies on one of the boundaries.
Since we will be comparing only networks of the same order, we will focus on the sum of inverse distances among the vertices of each network.
\begin{equation}
\mathcal{F}'(\mathcal{G})={N \choose 2} \mathcal{F}(\mathcal{G}).
\end{equation}
If a network $\mathcal{R}$ is not a minimum average distance graph, then according to Corollary \ref{NecessaryConditionForNonneighbors} there exists at least one pair of vertices $(a,b)$ with $d(a,b) \geq 3$.
The sum of the inverse shortest path lengths of such a network is 
\begin{equation}
\begin{aligned}
\mathcal{F}'(\mathcal{R}) &=\sum _{\substack{(u,v)\in \mathcal{V}^{2}(\mathcal{G})\\ u\neq v} } \frac{1}{h_{uv}} =\sum _{k \geq 1} \frac{1}{k} |\mathcal{E}_{k}(\mathcal{G})|\\
&=m +\frac{1}{2}|\mathcal{E}_{2}(\mathcal{R})|+ \sum _{k \geq 3} \frac{1}{k} |\mathcal{E}_{k}(\mathcal{R})|.
\end{aligned}
\end{equation}
On the other hand, the sum of the inverse distances of a minimum average distance network $\mathcal{G}$ is
\begin{equation}
\begin{aligned}
\mathcal{F}'(\mathcal{G}) &=\sum _{\substack{(u,v)\in \mathcal{V}^{2}(\mathcal{G})\\ u\neq v} } \frac{1}{d_{uv}}= m +\frac{1}{2}|\mathcal{E}_{2}(\mathcal{G})| \\
&=m +\frac{1}{2} \left( {N\choose 2} -m \right) =\frac{1}{2}m +\frac{1}{2} {N\choose 2}.
\end{aligned}
\end{equation}
The difference is therefore
\begin{equation}
\begin{aligned}
\mathcal{F}'(\mathcal{G}) -\mathcal{F}'(\mathcal{R}) &= \left( \frac{1}{2}m +\frac{1}{2} {N\choose 2} \right) -\left( m +\frac{1}{2}|\mathcal{E}_{2}(\mathcal{R})|+ \sum _{k \geq 3} \frac{1}{k} |\mathcal{E}_{k}(\mathcal{R})|  \right) \\
& \geq \frac{1}{2} {N\choose 2}-\frac{1}{2}m - \frac{1}{2}|\mathcal{E}_{2}(\mathcal{R})| - \frac{1}{3} \sum _{k \geq 3} |\mathcal{E}_{k}(\mathcal{R})| \\
&= \frac{1}{2}m + \frac{1}{2} {N\choose 2} - \frac{1}{2}|\mathcal{E}_{2}(\mathcal{R})| - \frac{1}{3} \left( {N\choose 2} -m -|\mathcal{E}_{2}(\mathcal{R})|  \right) \\
&= \frac{1}{6} \left( {N\choose 2} -m - |\mathcal{E}_{2}(\mathcal{R})|  \right) \\
&= \frac{1}{6} \sum _{k\geq 3}|\mathcal{E}_{k}(\mathcal{R})| \\
&>0.
 \end{aligned}
\end{equation}
This shows that a maximum efficiency graph is a minimum distance graph.
Normalizing by the total number of vertex pairs, equation \eqref{MaxEfficiencyEquation} follows.
\end{proof}

\begin{theorem}
A network has the lowest possible efficiency if and only if it is a largest average distance graph.
\end{theorem}
\begin{proof}
We will use the same method as in the proof for the form of networks with the largest average distance.
When ${N-1 \choose 2}+1 \leq m \leq {N \choose 2}$, then all networks have $m$ pairs of connected vertices, ${N \choose 2}-m$ pairs of vertices that are second neighbors, and there is no graph in which two vertices do not have any common neighbors, as shown in Lemma \ref{AlmostCompleteGraphsDistanceLemma}.
This clearly shows that all networks of this size have the same efficiency, given by equation \eqref{MaxEfficiencyEquation}.
For smaller size graphs, when $N-1 \leq m \leq {N-1 \choose 2}$, a necessary and sufficient condition will be
\begin{equation}
\mathcal{G} \in \mathcal{I} \iff \mathcal{G-H} \in \mathcal{I}  \qquad \forall \quad \mathcal{H} \subseteq \mathcal{G}.
\label{EverySubsetSmallestEfficiency}
\end{equation}
with $\mathcal{I}$ being the set of networks with the lowest efficiency.
If we consider a subgraph of order $1$ (a single vertex), its average distance to all other vertices will be the largest when its degree is equal to $1$.
So, if $\mathcal{G}=\mathcal{H} \cup \{u\}$, with $\mathcal{H} \in \mathcal{I}$ and $u$ is only connected to one other vertex in the graph (its distance to which is equal to $1$), it is evident that it has to be connected to one of the vertices with the largest average distance, which at the same time has the largest eccentricity.
\begin{equation}
\begin{aligned}
\mathcal{F}(\mathcal{G}) &=\mathcal{F}(\mathcal{H})+ \sum _{\substack{k \in \mathcal{V}(\mathcal{G}) \\ k \neq u  }} \frac{1}{d_{ku}} \\
&=\mathcal{F}(\mathcal{H})+ 1+\sum _{\substack{k \in \mathcal{V}(\mathcal{G}) \\ k \neq u, k\neq v  }} \frac{1}{1+d_{vu}}. \\
\end{aligned}
\end{equation}
The last equation shows that if $v$ is the vertex of $\mathcal{H}$ with degree $d_{v}=1$, then the new graph has the smallest possible efficiency.
\end{proof}
\section{Radius and Diameter}

The radius and the diameter of a graph are also measures that have to do with distance.
In order to define the radius of a graph, we need to find a central vertex in the network, the one that is the closest to all other vertices.
A network may have more than one central vertex.
We are often interested in the radius of a network when information is aggregated and distributed from a vertex high in the hierarchy to other vertices lower in the hierarchy.
The importance of a node is correlated with how central it is.
Important vertices are usually the ones closest to the network center.

On the other hand, the diameter of a network becomes important when we have a flat hierarchy, where communication or signal propagation takes place with the same frequency among any given pair of vertices in the network.
There are applications in which we want our network to have very small or very large diameter.
Usually for signal propagation or in general diffusion phenomena, the desired network architecture has the smallest possible diameter, since the response to different inputs needs to be processed as fast as possible.
When considering a virus spreading in the network during a fixed time interval, in order to ensure that as few nodes as possible get infected before appropriate action is taken, the network diameter has to be as large as possible.

Here, we are going to show the structure of the networks with the largest and smallest radius and diameter.
As we will see below, these graphs do not always have the same form.

\subsection{Networks with the Smallest and Largest Radius}

In this section, we will find tight bounds for the radius of graphs of arbitrary order and size.
The networks that achieve these bounds are not generally unique.
The radius of a network is correlated with its average distance and diameter.
Graphs with the smallest radius have the smallest average distance and smallest diameter, whereas graphs with the largest radius may or may not have the largest average distance or diameter, as we will see next.

\begin{lemma}
If $(u,v) \in \mathcal{E}(\mathcal{G})$, then 
\begin{equation}
ecc(u)-1 \leq ecc(v) \leq ecc(u)+1.
\end{equation}
\label{NeighborEccentricity}
\end{lemma}
\begin{proof}
For every vertices $u,v$ such that $(u,v) \in \mathcal{E}(\mathcal{G})$, and $w \in \mathcal{V}(\mathcal{G})$, 
\begin{equation}
\begin{aligned}
&|d(v,w)-d(u,w)| \leq 1 \\
&\implies d(u,w)-1 \leq d(v,w) \leq d(u,w)+1 \\
&\implies  \displaystyle \max _{w\in \mathcal{V}(\mathcal{G})}\{d(u,w)-1\} \leq \max _{w\in \mathcal{V}(\mathcal{G})} d(v,w) \leq \max _{w\in \mathcal{V}(\mathcal{G})} \{d(u,w)+1\} \\
&\implies ecc(u)-1 \leq ecc(v) \leq ecc(u)+1.
\end{aligned}
\end{equation}
\end{proof}

\begin{theorem}
A network of order $N$ and size $m$ has the smallest possible radius if and only if it has an induced subgraph which is the star graph. 
Such a network has a radius equal to one, regardless of its size.
\end{theorem}
\begin{proof}
The radius $R(\mathcal{G})$ of any graph $\mathcal{G}$ is a natural number, with $R(\mathcal{G})\geq 1$.
If a star of the same order as $\mathcal{G}$ is an induced subgraph, then the central vertex has eccentricity equal to one, which is the minimum possible.
Conversely, if the radius is equal to one, then there exists at least one vertex with full degree, which, along with its neighbors forms a star subgraph.
\end{proof}

\begin{corollary}
A network with the smallest radius also has the smallest average distance among its vertices.
The opposite is not necessarily true, since there exist minimum average distance networks with no vertices of full degree (Corollaries \ref{MinDistanceNecessaryCondition1}-\ref{MinDistanceNecessaryCondition3}).
\end{corollary}

\begin{lemma}
The maximum radius of a graph $\mathcal{G}(N,m)$ is a nonincreasing function with respect to the size $m$.
\label{RadiusDecreasingFunctionLemma}
\end{lemma}
\begin{proof}
Adding an edge to any graph $\mathcal{G}(N,m)$ will create a shorter path between at least two vertices, so the eccentricity of every vertex in $\mathcal{G}$ is either unchanged or decreases.
\end{proof}

\begin{lemma}
Assume that $\mathcal{G}(N,m)$ has radius $R(\mathcal{G})=r$,  and $c$ is a central vertex.
If $d(a,c)=r$ for some $a \in \mathcal{V}(\mathcal{G})$, then there exists a vertex $b \in \mathcal{V}(\mathcal{G})$ such that 
\begin{equation}
d(b,c) \geq r-1 \quad \textrm{and} \quad \mathcal{P}(a,c) \cap \mathcal{P}(b,c) =c.
\end{equation}
\label{GraphCenterCyclicProperty}
\end{lemma}
\begin{proof}
Suppose that there does not exist such a vertex.
If the first condition is not satisfied, then
\begin{equation}
d(u,c)\leq r-2 \quad \forall u \in \mathcal{V}(\mathcal{G}) \implies R(\mathcal{G}) \leq r-1
\end{equation}
which contradicts the assumption that $\mathcal{G}$ has radius $r$.

If there do not exist any vertices $a$ and $b$ with distances at least $r$ and $r-1$ respectively from $c$ whose shortest paths to $c$ have no other common vertex, then there exists a different vertex $w\in \mathcal{P}(a,c) \cap \mathcal{P}(b,c)$ such that
\begin{equation}
d(a,w)\leq r-1 \quad \textrm{and} \quad d(b,w)\leq r-2
\end{equation}
meaning that $c$ is not a central vertex.
\end{proof}

\begin{lemma}
A path graph has a radius larger or equal to any other tree network,
\begin{equation}
R_{max}(N,m=N-1)=\Big \lfloor \frac{N}{2} \Big \rfloor.
\end{equation}
A cycle graph has radius larger or equal to any other network, $R_{max}(N,m=N)=R_{max}(N,m=N-1)$.
\label{TreesAndCyclesMaxRadius}
\end{lemma}
\begin{proof}
A network $\mathcal{G}$ with radius $R \geq \big\lfloor \frac{N}{2} \big\rfloor+1$, according to Lemma \ref{GraphCenterCyclicProperty} will need to have an order of 
\begin{equation}
|\mathcal{V}(\mathcal{G})| \geq 1+1+\Big\lfloor \frac{N}{2} \Big\rfloor+\Big\lfloor \frac{N}{2} \Big\rfloor \geq N+1
\end{equation}
which is a contradiction.
If the path graph has an odd number of vertices, the central vertex is the middle vertex, with distance $\frac{N-1}{2}$ from both extreme vertices.
If the order is even, then both middle vertices are graph centers, and their eccentricities is equal to $\frac{N}{2}$.
Connecting the two vertices that are furthest from the center through an edge does not have an impact to the graph radius, so a cycle has the largest possible radius (Lemma \ref{RadiusDecreasingFunctionLemma}).
Because of the symmetry of the network, all vertices have the same eccentricity.
\end{proof}

\begin{lemma}
A graph of order $N$ and size ${N \choose 2}- \big\lceil \frac{N}{2}\big\rceil +1 \leq m \leq {N \choose 2} $ has radius equal to $1$.
\end{lemma}
\begin{proof}
It suffices to prove that there exists at least one vertex with full degree.
If all vertices have degree less than $N-2$, the graph size is at most
\begin{equation}
m \leq m_{0} =\Bigg\lfloor \frac{N(N-2)}{2} \Bigg \rfloor
\end{equation}
which is not possible, since for all $N\geq 2$
\begin{equation}
{N \choose 2}- \Bigg\lceil \frac{N}{2}\Bigg\rceil +1>m_{0}.
\end{equation}
\end{proof}

\begin{lemma}
The largest possible radius for a graph of order $N$ and size ${N-2 \choose 2}+1 \leq m \leq {N \choose 2}- \Big\lceil \frac{N}{2} \Big\rceil$ is equal to $2$.
\end{lemma}
\begin{proof}
It suffices to prove that graphs with size  $m\geq{N-2 \choose 2}+1$ cannot have radius of $3$ or larger, since we can find at least one network of this size in which no vertex has full degree \cite{DegreeRealizability}.
Refering to Figure \ref{Radius3NecessarySubgraph},  let $C$ be one of the central vertices.
According to Lemma \ref{GraphCenterCyclicProperty}, there exist at least two vertices $A$ and $F$ (possibly connected to each other) such that
\begin{equation}
d(C,A)\geq 2 \quad \textrm{and} \quad d(C,F)\geq 3.
\end{equation}
In order to respect these distance conditions and the centrality of $C$,
\begin{equation}
C\notslash F,\quad B\notslash E \quad\textrm{and}\quad A\notslash D.
\end{equation}
Because $d(C,F)\geq 3$, nodes $C$ and $F$ cannot have a common neighbor, so there are $N-2$ edges that are \textit{not} present in the graph.
In addition, $B$ and $E$ may not have a common neighbor either, otherwise the radius would be at most equal to $2$.
There are another $N-4$ edges that cannot be present, since there are $N-2$ possible common neighbors of $B$ and $E$, and we have already counted two of them in the previous case.
So the graph is missing at least $m_{s}=(N-2)+(N-4)+3$ edges, and its size is at most
\begin{equation}
\begin{aligned}
m&\leq {N \choose 2} -(2N-3) \\
&={N-2 \choose 2}.
\end{aligned}
\end{equation}
\end{proof}

\begin{center}
\begin{figure}[htb]
\subfigure[]{
\psscalebox{0.4}{
\begin{pspicture}(0,-4)(10,1)
{
\cnodeput(0,0){A}{\strut\boldmath$A$}
\cnodeput(2,0){B}{\strut\boldmath$B$}
\cnodeput(4,0){C}{\strut\boldmath$C$}
\cnodeput(6,0){D}{\strut\boldmath$D$}
\cnodeput(8,0){E}{\strut\boldmath$E$}
\cnodeput(10,0){F}{\strut\boldmath$F$}
}
\ncline{-}{A}{B}
\ncline{-}{B}{C}
\ncline{-}{C}{D}
\ncline{-}{D}{E}
\ncline{-}{E}{F}
\ncline{-}{F}{G}
\ncarc[linestyle=dashed, linewidth=1pt,arcangle=-40]{A}{F}
\end{pspicture}
}
\label{Radius3NecessarySubgraph}
}
\subfigure[]{
\psscalebox{0.4}{
\begin{pspicture}(-1,-4)(13.5,1)
{
\cnodeput(0,0){A}{\strut\boldmath$A$}
\cnodeput(4,0){B}{\strut\boldmath$B$}
\cnodeput(6,0){C}{\strut\boldmath$C$}
\cnodeput(8,0){D}{\strut\boldmath$D$}
\cnodeput(12,0){F}{\strut\boldmath$F$}
\cnodeput(4.5,-1.7){G}{\strut\boldmath$G$}
\cnodeput(7.5,-1.7){H}{\strut\boldmath$H$}
\cnodeput(6,-3){K}{\strut}
}
\ncline{-}{B}{C}
\ncline{-}{C}{D}
\ncline{-}{B}{G}
\ncline{-}{D}{H}
\ncline{-}{C}{G}
\ncline{-}{C}{H}
\ncline{-}{C}{K}
\ncline{-}{G}{H}
\ncline{-}{B}{H}
\ncline{-}{D}{G}
\nccoil[coilaspect=0,coilheight=2.5,coilwidth=0.2,coilarm=0.2cm]{-}{A}{B}
\nccoil[coilaspect=0,coilheight=2.5,coilwidth=0.2,coilarm=0.2cm]{-}{D}{F}
\ncarc[linestyle=dashed, linewidth=1pt,arcangle=40]{A}{F}
\ncarc[linestyle=dashed, linewidth=1pt,arcangle=-30]{G}{K}
\ncarc[linestyle=dashed, linewidth=1pt,arcangle=30]{H}{K}
\end{pspicture}
}
\label{CycleSufficientForMaxRadius}
}
\subfigure[]{
\psscalebox{0.5}{
\begin{pspicture}(-2,-2.5)(2.2,2)
{
\cnodeput(1.24,1.56){A}{}
\cnodeput[fillstyle=solid,fillcolor=red](-0.45,1.95){B}{}
\cnodeput(-1.8,0.87){C}{}
\cnodeput(-1.8,-0.87){D}{}
\cnodeput(-0.45,-1.95){E}{}
\cnodeput(1.25,-1.56){F}{}
\cnodeput(2,0){G}{}
}
\ncline{-}{A}{B}
\ncline{-}{B}{C}
\ncline{-}{C}{D}
\ncline{-}{D}{E}
\ncline{-}{E}{F}
\ncline{-}{F}{G}
\ncline{-}{G}{A}
\ncline{-}{A}{C}
\ncline{-}{A}{D}
\ncline{-}{B}{C}
\ncline{-}{B}{D}
\ncline{-}{B}{G}
\end{pspicture}
}
\label{GenericMaxRadiusExample}
}
\caption{\textbf{(a)} Necessary induced subgraph for a network to have radius larger or equal to $3$. The vertices shown are not allowed to have any direct connections, and vertices with distance $3$ are not allowed to have any common neighbors. \textbf{(b)} For order $N$ and size $m\geq N$, there is always at least one graph $\mathcal{G}(N,m)$ that has the maximum possible radius, and has a full cycle as an induced subgraph. \textbf{(c)} A network with largest possible radius with $N=7$ and $m=11$. It consists of an almost complete subgraph of order $M=5$ and a path graph of order $L=4$. The two induced subgraphs share two vertices, so that $N=L+M-2$. The red vertex has the largest degree, and is the center of the network.}
\end{figure}
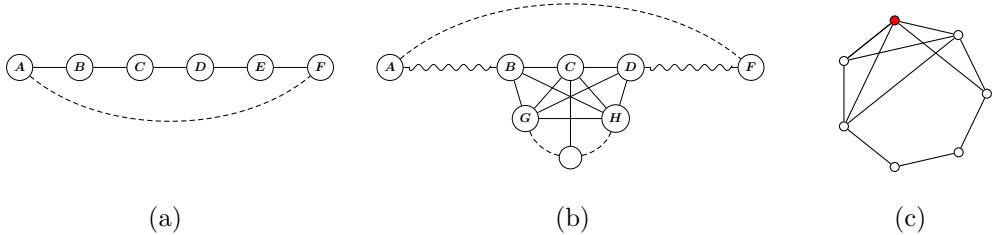
\end{center}

\begin{lemma}
The maximum possible radius of a graph of order $N$ and size $N+1\leq m \leq {N-2 \choose 2}-2$ is
\begin{equation}
R_{max}(N,m)=\Bigg\lceil \frac{2N-1-\sqrt{1+8m-8N}}{4}\Bigg\rceil.
\end{equation}
\end{lemma}
\begin{proof}
For every $N\geq 3$, there is at least one graph with radius equal to $R_{max}(N,m)$ that includes a full circle as an induced subgraph.
To see why, suppose that $C$ is a central vertex and 
\begin{equation}
d(C,A)=R_{max}\quad\textrm{and}\quad d(C,F)\geq R_{max}-1 \qquad A,F \in \mathcal{V}(\mathcal{G})
\label{FullCycleDistanceFromCentralVertex}
\end{equation}
as shown in Figure \ref{CycleSufficientForMaxRadius}.
We pick vertices $B$ and $D$ such that
\begin{equation}
d(C,B)=1 \quad d(C,D)=1, \textrm{ and } B \in \mathcal{P}(C,A),  G \in \mathcal{P}(C,F).
\end{equation}
Also, all other vertices can be connected to $B,D$ and $C$ without changing the radius, when $m \leq {N-2 \choose 2}-2$.
Vertices $A$ and $F$ can be connected without changing the maximum radius of the network, as shown in Lemma \ref{TreesAndCyclesMaxRadius}.
Also, there are no edges among vertices in $\mathcal{P}(C,A)$ or vertices that belong to $\mathcal{P}(C,A)$, otherwise condition \eqref{FullCycleDistanceFromCentralVertex} would not be satisfied.
Thus, the network described has a full circle as an induced subgraph.

We now need to compute the maximum radius of such a graph.
Since a new edge always creates new shortest paths, we have to connect vertices with distance equal to two, such that we only create a single new shortest path between vertices that are second neighbors (see also equation \eqref{RewiringToSecondNeighbors}).
In other words, a simple method to find the graph with the largest possible radius is to start from a cycle graph, and keep adding edges such that we have a complete or almost complete graph connected with both ends of a path graph, as shown in Figure \ref{GenericMaxRadiusExample}.
This process is the same as the one for finding networks that minimize crosstalk among individual elements \cite{MinXtalkNetworks}, with the only difference being that the initial graph is a cycle instead of a star.
Based on the symmetry of this type of network, we can assume that the vertex with the largest degree is always a central vertex.
If we denote with $M$ the order of the complete or almost complete graph, and with $L$ the order of the path graph, we can find these orders by solving the following system of equations:
\begin{equation}
\begin{aligned}
m&=(L-1)+{M-1 \choose 2}+\alpha  \\
L+M &=N+2
\end{aligned}
\end{equation}
with
\begin{equation}
1 \leq \alpha \leq M-1, \quad L,C \in \mathbb{N}^{*}.
\end{equation}
We compute $M$ (and subsequently $L$) in the same way as in the proof of Corollary \ref{MaxDistanceCalculation}, by setting $\alpha$ equal to its maximum value, and then choosing the smallest integer $M$ that is smaller than the solution of the second order equation.
\begin{equation}
M=\Bigg \lfloor \frac{3+\sqrt{1+8m-8N}}{2}\Bigg\rfloor, \quad  L=N+2-\Bigg \lfloor \frac{3+\sqrt{1+8m-8N}}{2} \Bigg\rfloor.
\end{equation}
This graph has radius equal to 
\begin{equation}
\begin{aligned}
R_{max}(N,m)&=\Bigg\lfloor \frac{L+\delta(a<M-1)}{2}\Bigg\rfloor \\
&=\Bigg\lfloor \frac{N+2+\delta(a<M-1) -\big \lfloor \frac{3+\sqrt{1+8m-8N}}{2} \big\rfloor}{2}\Bigg\rfloor 
\end{aligned}
\end{equation}
where $\delta$ is the Kronecker delta function.
After simplifying, the last expression becomes:
\begin{equation}
R_{max}(N,m)=\Bigg\lceil \frac{2N-1-\sqrt{1+8m-8N}}{4}\Bigg\rceil.
\end{equation}
\end{proof}

\begin{theorem}
The maximum radius of a network of order $N$ and size $m$ is
\begin{equation}
R_{max}(N,m)= \left\{ \begin{array}{ll}
\lfloor \frac{N}{2}  \rfloor & N-1\leq m \leq N\\
\Big\lceil \frac{2N-1-\sqrt{1+8m-8N}}{4}\Big\rceil & N+1 \leq m \leq {N-2 \choose 2}\\
2 &  {N-2 \choose 2}+1 \leq m \leq {N \choose 2}- \big\lceil \frac{N}{2} \big\rceil \\
1 & {N \choose 2}- \big\lceil \frac{N}{2}\big\rceil +1 \leq m \leq {N \choose 2}.
\end{array} \right.
\end{equation}
\end{theorem}

\begin{figure}[htb]
\begin{center}
\includegraphics[scale=0.37]{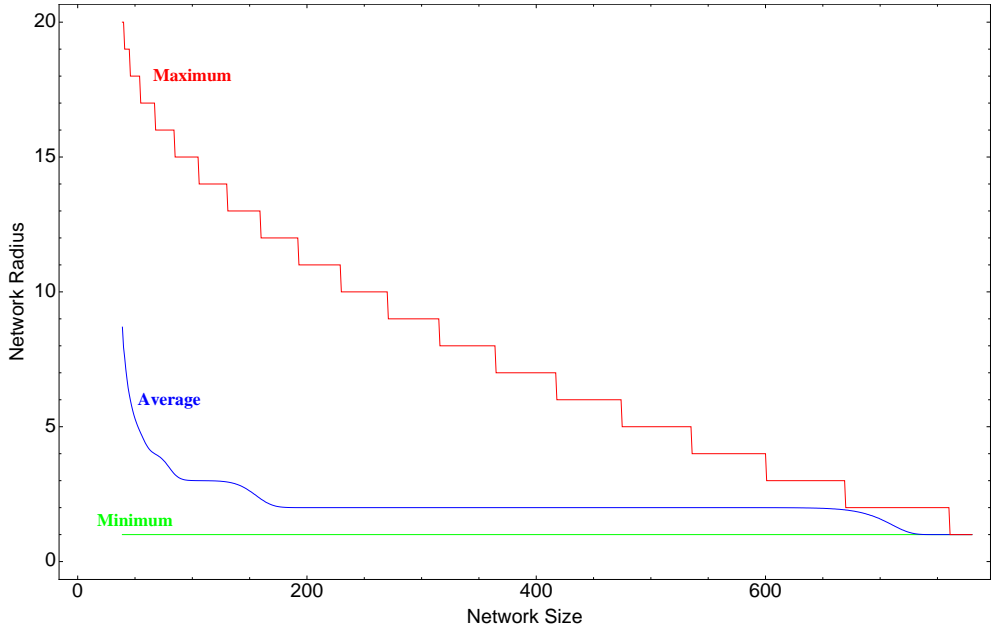}
\caption{The largest, smallest and average radius of graphs of order $N=40$ as a function their size $39\leq m \leq 780$. The minimum and maximum bounds are analytically computed. The statistical averages are estimated by the mean radius of $10^{4}$ random graphs for each size. The average radius also has a ``stepwise'' form.}
\label{MinAvgMaxRadiusBoundsExample}
\end{center}
\end{figure}

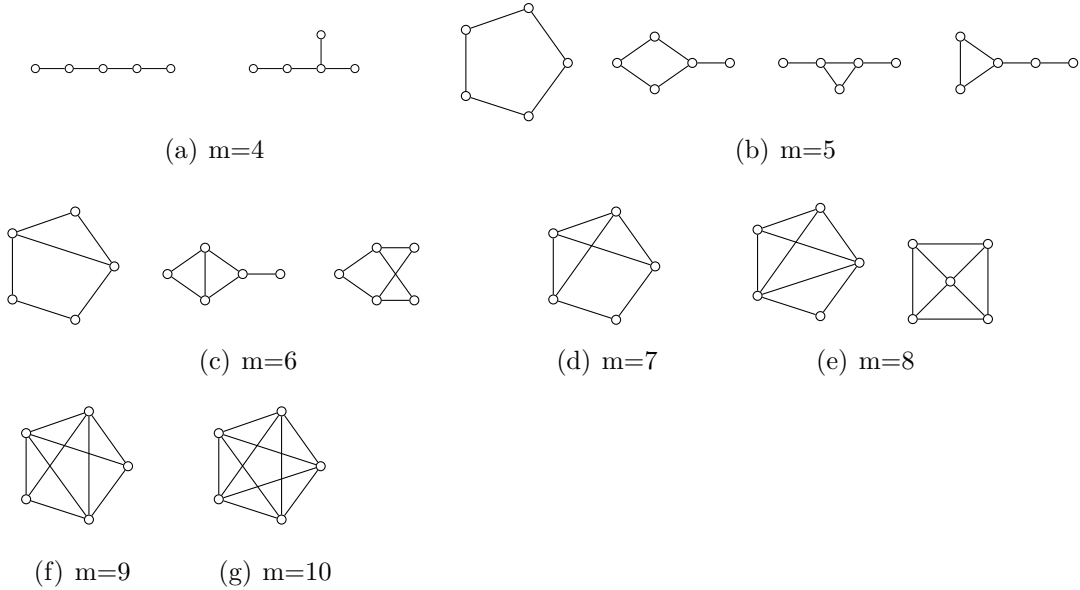
\begin{figure}[htb]
\subfigure[m=4]{
\psscalebox{0.45}{
\begin{pspicture}(-1,-1.5)(5,1)
{
\cnodeput(0,0){A}{}
\cnodeput(1,0){B}{}
\cnodeput(2,0){C}{}
\cnodeput(3,0){D}{}
\cnodeput(4,0){E}{}
}
\ncline{-}{A}{B}
\ncline{-}{B}{C}
\ncline{-}{C}{D}
\ncline{-}{D}{E}
\end{pspicture}
}
\psscalebox{0.45}{
\begin{pspicture}(-1,-1.5)(5,1)
{
\cnodeput(0,0){A}{}
\cnodeput(1,0){B}{}
\cnodeput(2,0){C}{}
\cnodeput(3,0){D}{}
\cnodeput(2,1){E}{}
}
\ncline{-}{A}{B}
\ncline{-}{B}{C}
\ncline{-}{C}{D}
\ncline{-}{C}{E}
\end{pspicture}
}
}
\subfigure[m=5]{
\psscalebox{0.5}{
\begin{pspicture}(-1.5,-1.5)(2.2,1.5)
{
\cnodeput(.46,1.46){A}{}
\cnodeput(-1.21,0.88){B}{}
\cnodeput(-1.21,-0.88){C}{}
\cnodeput(0.46,-1.42){D}{}
\cnodeput(1.5,0){E}{}
}
\ncline{-}{A}{B}
\ncline{-}{B}{C}
\ncline{-}{C}{D}
\ncline{-}{D}{E}
\ncline{-}{E}{A}
\end{pspicture}
}
\psscalebox{0.5}{
\begin{pspicture}(-.2,-1.5)(4,1)
{
\cnodeput(0,0){A}{}
\cnodeput(1,-0.7){B}{}
\cnodeput(1,0.7){C}{}
\cnodeput(2,0){D}{}
\cnodeput(3,0){E}{}
}
\ncline{-}{A}{B}
\ncline{-}{A}{C}
\ncline{-}{B}{D}
\ncline{-}{C}{D}
\ncline{-}{D}{E}
\end{pspicture}
}
\psscalebox{0.5}{
\begin{pspicture}(0,-1.5)(4,1)
{
\cnodeput(0,0){A}{}
\cnodeput(1,0){B}{}
\cnodeput(1.5,-0.7){C}{}
\cnodeput(2,0){D}{}
\cnodeput(3,0){E}{}
}
\ncline{-}{A}{B}
\ncline{-}{B}{C}
\ncline{-}{B}{D}
\ncline{-}{C}{D}
\ncline{-}{D}{E}
\end{pspicture}
}
\psscalebox{0.5}{
\begin{pspicture}(-0.3,-1.5)(4,1)
{
\cnodeput(0,-0.7){A}{}
\cnodeput(0,0.7){B}{}
\cnodeput(1,0){C}{}
\cnodeput(2,0){D}{}
\cnodeput(3,0){E}{}
}
\ncline{-}{A}{B}
\ncline{-}{A}{C}
\ncline{-}{B}{C}
\ncline{-}{C}{D}
\ncline{-}{D}{E}
\end{pspicture}
}
}
\subfigure[m=6]{
\psscalebox{0.5}{
\begin{pspicture}(-1.5,-1.7)(1.5,2)
{
\cnodeput(.46,1.46){A}{}
\cnodeput(-1.21,0.88){B}{}
\cnodeput(-1.21,-0.88){C}{}
\cnodeput(0.46,-1.42){D}{}
\cnodeput(1.5,0){E}{}
}
\ncline{-}{A}{B}
\ncline{-}{B}{C}
\ncline{-}{C}{D}
\ncline{-}{D}{E}
\ncline{-}{E}{A}
\ncline{-}{B}{E}
\end{pspicture}
}
\psscalebox{0.5}{
\begin{pspicture}(-1,-1.5)(4,2.5)
{
\cnodeput(0,0){A}{}
\cnodeput(1,-0.7){B}{}
\cnodeput(1,0.7){C}{}
\cnodeput(2,0){D}{}
\cnodeput(3,0){E}{}
}
\ncline{-}{A}{B}
\ncline{-}{A}{C}
\ncline{-}{B}{C}
\ncline{-}{B}{D}
\ncline{-}{C}{D}
\ncline{-}{D}{E}
\end{pspicture}
}
\psscalebox{0.5}{
\begin{pspicture}(-0.15,-1.5)(4,2.5)
{
\cnodeput(0,0){A}{}
\cnodeput(1,-0.7){B}{}
\cnodeput(1,0.7){C}{}
\cnodeput(2,-0.7){D}{}
\cnodeput(2,0.7){E}{}
}
\ncline{-}{A}{B}
\ncline{-}{A}{C}
\ncline{-}{B}{D}
\ncline{-}{B}{E}
\ncline{-}{C}{D}
\ncline{-}{C}{E}
\end{pspicture}
}
}
\subfigure[m=7]{
\psscalebox{0.5}{
\begin{pspicture}(-2,-1.7)(2.3,2)
{
\cnodeput(.46,1.46){A}{}
\cnodeput(-1.21,0.88){B}{}
\cnodeput(-1.21,-0.88){C}{}
\cnodeput(0.46,-1.42){D}{}
\cnodeput(1.5,0){E}{}
}
\ncline{-}{A}{B}
\ncline{-}{B}{C}
\ncline{-}{C}{D}
\ncline{-}{D}{E}
\ncline{-}{E}{A}
\ncline{-}{B}{E}
\ncline{-}{A}{C}
\end{pspicture}
}
}
\subfigure[m=8]{
\psscalebox{0.5}{
\begin{pspicture}(-2.5,-1.8)(2,2)
{
\cnodeput(.46,1.46){A}{}
\cnodeput(-1.21,0.88){B}{}
\cnodeput(-1.21,-0.88){C}{}
\cnodeput(0.46,-1.42){D}{}
\cnodeput(1.5,0){E}{}
}
\ncline{-}{A}{B}
\ncline{-}{B}{C}
\ncline{-}{C}{D}
\ncline{-}{D}{E}
\ncline{-}{E}{A}
\ncline{-}{B}{E}
\ncline{-}{A}{C}
\ncline{-}{C}{E}
\end{pspicture}
}
\psscalebox{0.5}{
\begin{pspicture}(-1.5,-1.3)(1.5,1.5)
{
\cnodeput(-1,-1){A}{}
\cnodeput(-1,1){B}{}
\cnodeput(1,1){C}{}
\cnodeput(1,-1){D}{}
\cnodeput(0,0){E}{}
}
\ncline{-}{A}{B}
\ncline{-}{B}{C}
\ncline{-}{C}{D}
\ncline{-}{D}{A}
\ncline{-}{A}{E}
\ncline{-}{B}{E}
\ncline{-}{C}{E}
\ncline{-}{D}{E}
\end{pspicture}
}
}
\subfigure[m=9]{
\psscalebox{0.5}{
\begin{pspicture}(-2,-2)(2.3,2)
{
\cnodeput(.46,1.46){A}{}
\cnodeput(-1.21,0.88){B}{}
\cnodeput(-1.21,-0.88){C}{}
\cnodeput(0.46,-1.42){D}{}
\cnodeput(1.5,0){E}{}
}
\ncline{-}{A}{B}
\ncline{-}{A}{C}
\ncline{-}{A}{D}
\ncline{-}{A}{E}
\ncline{-}{B}{C}
\ncline{-}{B}{D}
\ncline{-}{B}{E}
\ncline{-}{C}{D}
\ncline{-}{D}{E}
\end{pspicture}
}
}
\subfigure[m=10]{
\psscalebox{0.5}{
\begin{pspicture}(-2,-2)(2.3,2)
{
\cnodeput(.46,1.46){A}{}
\cnodeput(-1.21,0.88){B}{}
\cnodeput(-1.21,-0.88){C}{}
\cnodeput(0.46,-1.42){D}{}
\cnodeput(1.5,0){E}{}
}
\ncline{-}{A}{B}
\ncline{-}{A}{C}
\ncline{-}{A}{D}
\ncline{-}{A}{E}
\ncline{-}{B}{C}
\ncline{-}{B}{D}
\ncline{-}{B}{E}
\ncline{-}{C}{D}
\ncline{-}{C}{E}
\ncline{-}{D}{E}
\end{pspicture}
}
}
\caption{All connected networks with $5$ vertices and $4\leq m \leq 10$ edges with the largest possible radius. The first graph of each group is constructed by the method described in the text.}
\label{MaxRadius5VerticesAll}
\end{figure}

An example of the form of the function above for $N=40$ and all sizes is shown in Figure \ref{MinAvgMaxRadiusBoundsExample}.
Note the same stair-like form of both the maximum radius and the statistical averages.
The statistical average curve exhibits fewer and smoother ``steps''. 
Networks with the largest radius of order $N=5$ and all sizes are listed in Figure \ref{MaxRadius5VerticesAll}.

\FloatBarrier


\subsection{Networks with the Smallest and Largest Diameter}

In this subsection, we are going to study the form of the networks with the minimum and maximum diameter.
Computing the minimum diameter of a network is fairly straightforward.
In the case of the maximum possible diameter, we first prove two lemmas that will help us show that we can find the structure of the networks recursively.

\begin{theorem}
A network has the smallest possible diameter if and only if it is a smallest average distance graph.
\end{theorem}
\begin{proof}
The diameter of a complete graph is trivially equal to one.
If the graph is not complete, the diameter is at least $2$, since there is at least one pair of non-neighboring vertices.
In a graph with the smallest average distance, all vertices that are not connected have at least one common neighbor, and the maximum eccentricity is equal to $2$.
Conversely, if the largest distance among any vertex pair is equal to $2$, then by Corollary \ref{NecessaryConditionForNonneighbors}, the graph has the smallest average distance.
\end{proof}

\begin{corollary}
A network with the minimum radius ($R(\mathcal{G})=1$) also has minimum diameter ($T(\mathcal{G})=2$) regardless of its interconnection topology.
The inverse is not always true: There are networks with minimum diameter, and radius $R(\mathcal{G})=2>R_{min}(N,m)$.
\end{corollary}

\begin{lemma}
A network of order $N$ and size ${N-1 \choose 2}+1 \leq m \leq {N \choose 2}-1$ has a diameter equal to $2$. A complete graph has diameter equal to $1$.
\end{lemma}
\begin{proof}
In a complete graph, all vertices are connected to each other, so the eccentricity of every vertex is trivially equal to $1$.
In a graph of size $m\geq {N-1 \choose 2}+1$, all vertices that do not share an edge have at least one common neighbor, as shown in the proof of Lemma \ref{AlmostCompleteGraphsDistanceLemma}.
Consequently, every vertex has eccentricity either $1$ or $2$, so the diameter is equal to $2$ regardless of the graph topology.
\end{proof}

\begin{lemma}
The largest possible diameter $T_{max}(N,m)$ of a network of order $N$ is at most one larger than the largest possible diameter of a network with order $N-1$ and smaller size.
\begin{equation}
T_{max}(N,m) \leq T_{max}(N-1,m-d)+1,\quad 1\leq d \leq N-1.
\end{equation}
\label{BasicInductionInequalityLemma}
\end{lemma}
\begin{proof}
Assume that the graph $\mathcal{G}=\mathcal{G}_{N,m}$ has diameter $T(\mathcal{G}) \leq T_{max}(N,m)$.
Define as $\mathcal{D}$ the set of unordered vertex pairs whose distance is equal to the graph diameter.
We now remove an arbitrary vertex $u$ with degree  $d=d_{u}$ from $\mathcal{G}$, and the resulting graph is $\mathcal{H}$ with order $N-1$.
If $u \in \mathcal{D}$, then no shortest path between any vertex pair in $\mathcal{D}$ passes through $u$.
We distinguish two cases:
\begin{itemize}
\item
If $u$ is in every vertex pair in $\mathcal{D}$, then the diameter of $\mathcal{H}$ is
\begin{equation}
\begin{aligned}
T(\mathcal{G}) &=T(\mathcal{H})+1 \qquad \forall \mathcal{G},\mathcal{H}\\
\implies T_{max}(N,m) &\leq T_{max}(N-1,m-d)+1.
\end{aligned}
\end{equation}
\item
If there exists at least one vertex pair in $\mathcal{D}$ that does not include $u$, then removing $u$ will result in a graph $\mathcal{H}$ that has the same diameter as $\mathcal{G}$.
\begin{equation}
\begin{aligned}
T(\mathcal{G})=T(\mathcal{H}) & \leq T_{max}(N-1,m-d) \\
\implies T_{max}(N,m) &\leq T_{max}(N-1,m-d).
\end{aligned}
\end{equation}
\end{itemize}
Combining the two cases, the result follows.
\end{proof}

\begin{corollary}
If we remove a vertex $u$ from a graph $\mathcal{G}$, resulting to graph $\mathcal{H}$, then
\begin{equation}
T_{\mathcal{H}}=T_{\mathcal{G}}-1 \implies \mathcal{D}_{\mathcal{G}}= \{(u,w_{1}),(u,w_{2}),\dots,(u,w_{d}) \}.
\end{equation}
Conversely, if we add a vertex $u$ with degree $d$ to $\mathcal{H}$, then
\begin{equation}
T_{\mathcal{G}}=T_{\mathcal{H}}+1 \implies \{(v_{1},w_{1}),(v_{2},w_{2}),\dots,(v_{d},w_{d}) \} \subseteq \mathcal{D}_{\mathcal{H}}
\end{equation}
with $w_{1},\dots, w_{d},v_{1},\dots v_{d} \in \mathcal{V}(\mathcal{H}), \mathcal{V}(\mathcal{G})$ and $(u,v_{1}),\dots, (u,v_{d}) \in \mathcal{E}(\mathcal{G})$.
\label{IncrementalLargerDiameterLemma}
\end{corollary}

\begin{theorem}
The largest possible diameter of a network of order $N$ and size $m$ is equal to 
\begin{equation}
T_{max}(N,m)=N- \Bigg\lfloor \frac{1+\sqrt{9+8m-8N}}{2} \Bigg\rfloor.
\end{equation}
\end{theorem}

\begin{figure}[htbp]
\begin{center}
\includegraphics[scale=0.37]{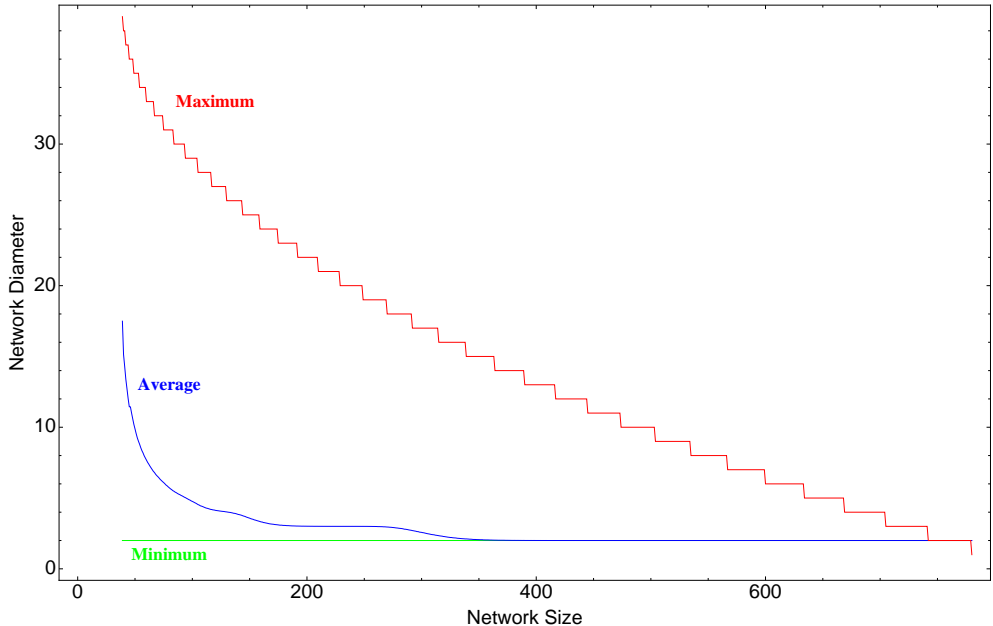}
\caption{The smallest, average and largest diameter of graphs of order $N=40$ as a function of their size, $39\leq m \leq 780$. The minimum and maximum bounds are analytically computed. The statistical average is approximated by the mean diameter of $10^{4}$ random graphs of the respective size. Note that even in the average diameter, there is the ``stepwise'' form of the maximum diameter graphs, although much smoother.}
\label{MinAvgMaxDiameterBoundsExample}
\end{center}
\end{figure}

\begin{proof}
Lemma \ref{BasicInductionInequalityLemma} and Corollary \ref{IncrementalLargerDiameterLemma} readily show an easy way to find the largest possible diameter of a graph of fixed order and size.
According to Corollary \ref{IncrementalLargerDiameterLemma}, if we add a vertex $u$ with degree $d_{u}=1$ to a maximum diameter graph $\mathcal{H}$, and we connect it to a vertex with the largest eccentricity, the resulting graph $\mathcal{G}$ has also the largest diameter for its order and size (see also Lemma \ref{NeighborEccentricity}).
As a result, we can write the following recursive relation:
\begin{equation}
T_{max}(N,m)=1+T_{max}(N-1,m-1).
\end{equation}
Repeating the process as many times as possible, 
\begin{equation}
T_{max}(N,m)=k+T_{max}(N-k,m-k).
\end{equation}
The only reason why we cannot continue the recursion relations further is that there cannot exist a network with $N-k-1$ vertices and $m-k-1$ edges, simply because
\begin{equation}
m-k-1>{N-k-1 \choose 2}.
\label{DiameterStoppingCondition}
\end{equation}
This means that the subgraph $\mathcal{H}$ with $N-k$ vertices and order $m-k$ has size
\begin{equation}
{N_{\mathcal{H}}-1 \choose 2} \leq m_{\mathcal{H}} \leq {N_{\mathcal{H}} \choose 2}
\end{equation}
 and consequently has a diameter of $2$, so the maximum diameter of the graph is:
\begin{equation}
T_{max}(N,m)=k+2
\label{MaxDiameterFinalRecursiveEquation}
\end{equation}
where $k=N-N_{\mathcal{H}}$.
One of the graphs with the maximum diameter consists of a path graph with $N-N_{\mathcal{H}}$ vertices, and a subgraph of order $N_{\mathcal{H}}$ and size $m_{\mathcal{H}} \geq {N_{\mathcal{H}}-1 \choose 2}+1$.
If we assume without loss of generality that $\mathcal{H}$ is a type $I$ almost complete graph, then $\mathcal{G}$ is a maximum distance network.
It consists of a path graph of order $P=k+1$ and a complete graph of order $C$, such that $P+C=N$, as shown in Figure \ref{MaxDistanceGeneralCase}.
If we denote by $C$ and $P$ the order of the clique and the path subgraphs as in equations \eqref{MaxDistanceGraphCliqueOrder} and \eqref{MaxDistanceGraphPathOrder}, and combine them with equation \eqref{MaxDiameterFinalRecursiveEquation}, the maximum diameter of a graph $\mathcal{G}(N,m)$ is
\begin{equation}
\begin{aligned}
T_{max}(N,m) &=k+2=P+1=N-C+1 \\
&=N- \Bigg\lfloor \frac{1+\sqrt{9+8m-8N}}{2} \Bigg\rfloor .\\
\end{aligned}
\end{equation} 
\end{proof}

\begin{corollary}
A maximum average distance graph also has the largest possible diameter. The converse is not always true.
\end{corollary}

An example of the minimum, maximum and average diameter of graphs with $40$ vertices and increasing number of edges is shown in Figure \ref{MinAvgMaxDiameterBoundsExample}.

\begin{figure}[htbp]
\subfigure[m=4]{
\psscalebox{0.45}{
\begin{pspicture}(-1,-1.5)(5,1)
{
\cnodeput(0,0){A}{}
\cnodeput(1,0){B}{}
\cnodeput(2,0){C}{}
\cnodeput(3,0){D}{}
\cnodeput(4,0){E}{}
}
\ncline{-}{A}{B}
\ncline{-}{B}{C}
\ncline{-}{C}{D}
\ncline{-}{D}{E}
\end{pspicture}
}
}
\subfigure[m=5]{
\psscalebox{0.5}{
\begin{pspicture}(-1,-1.5)(4,1)
{
\cnodeput(0,-0.7){A}{}
\cnodeput(0,0.7){B}{}
\cnodeput(1,0){C}{}
\cnodeput(2,0){D}{}
\cnodeput(3,0){E}{}
}
\ncline{-}{A}{B}
\ncline{-}{A}{C}
\ncline{-}{B}{C}
\ncline{-}{C}{D}
\ncline{-}{D}{E}
\end{pspicture}
}
\psscalebox{0.5}{
\begin{pspicture}(0,-1.5)(4,1)
{
\cnodeput(0,0){A}{}
\cnodeput(1,-0.7){B}{}
\cnodeput(1,0.7){C}{}
\cnodeput(2,0){D}{}
\cnodeput(3,0){E}{}
}
\ncline{-}{A}{B}
\ncline{-}{A}{C}
\ncline{-}{B}{D}
\ncline{-}{C}{D}
\ncline{-}{D}{E}
\end{pspicture}
}
\psscalebox{0.5}{
\begin{pspicture}(0,-1.5)(4,1)
{
\cnodeput(0,0){A}{}
\cnodeput(1,0){B}{}
\cnodeput(1.5,-0.7){C}{}
\cnodeput(2,0){D}{}
\cnodeput(3,0){E}{}
}
\ncline{-}{A}{B}
\ncline{-}{B}{C}
\ncline{-}{B}{D}
\ncline{-}{C}{D}
\ncline{-}{D}{E}
\end{pspicture}
}
}
\subfigure[m=6]{
\psscalebox{0.5}{
\begin{pspicture}(-1,-1.5)(4,2)
{
\cnodeput(0,0){A}{}
\cnodeput(1,-0.7){B}{}
\cnodeput(1,0.7){C}{}
\cnodeput(2,0){D}{}
\cnodeput(3,0){E}{}
}
\ncline{-}{A}{B}
\ncline{-}{A}{C}
\ncline{-}{B}{C}
\ncline{-}{B}{D}
\ncline{-}{C}{D}
\ncline{-}{D}{E}
\end{pspicture}
}
}
\subfigure[m=7]{
\psscalebox{0.5}{
\begin{pspicture}(0,-1)(4,2)
{
\cnodeput(0,0){A}{}
\cnodeput(1,-0.7){B}{}
\cnodeput(1,0.7){C}{}
\cnodeput(2,0){D}{}
\cnodeput(3,0){E}{}
}
\ncline{-}{A}{B}
\ncline{-}{A}{C}
\ncline{-}{A}{D}
\ncline{-}{B}{C}
\ncline{-}{B}{D}
\ncline{-}{C}{D}
\ncline{-}{D}{E}
\end{pspicture}
}
\psscalebox{0.5}{
\begin{pspicture}(0,-1)(4,2)
{
\cnodeput(0,0){A}{}
\cnodeput(1,-0.7){B}{}
\cnodeput(1,0.7){C}{}
\cnodeput(2,0){D}{}
\cnodeput(3,0){E}{}
}
\ncline{-}{A}{B}
\ncline{-}{A}{C}
\ncline{-}{B}{C}
\ncline{-}{B}{D}
\ncline{-}{C}{D}
\ncline{-}{B}{E}
\ncline{-}{C}{E}
\end{pspicture}
}
\psscalebox{0.5}{
\begin{pspicture}(0,-1)(5,2)
{
\cnodeput(0,-.5){A}{}
\cnodeput(1,.5){B}{}
\cnodeput(3,.5){C}{}
\cnodeput(2,-.5){D}{}
\cnodeput(4,-.5){E}{}
}
\ncline{-}{A}{D}
\ncline{-}{D}{E}
\ncline{-}{A}{B}
\ncline{-}{B}{C}
\ncline{-}{B}{D}
\ncline{-}{C}{D}
\ncline{-}{C}{E}
\end{pspicture}
}
\psscalebox{0.5}{
\begin{pspicture}(0,-1)(4,2)
{
\cnodeput(0,-.5){A}{}
\cnodeput(0,.5){B}{}
\cnodeput(1,-.5){C}{}
\cnodeput(1,.5){D}{}
\cnodeput(2,0){E}{}
}
\ncline{-}{A}{B}
\ncline{-}{A}{C}
\ncline{-}{A}{D}
\ncline{-}{B}{C}
\ncline{-}{B}{D}
\ncline{-}{C}{E}
\ncline{-}{D}{E}
\end{pspicture}
}}
\subfigure[m=8]{
\psscalebox{0.5}{
\begin{pspicture}(0,-0.5)(3,2)
{
\cnodeput(0,0){A}{}
\cnodeput(0,1){B}{}
\cnodeput(1,0){C}{}
\cnodeput(1,1){D}{}
\cnodeput(2,0.5){E}{}
}
\ncline{-}{A}{B}
\ncline{-}{A}{C}
\ncline{-}{A}{D}
\ncline{-}{B}{C}
\ncline{-}{B}{D}
\ncline{-}{C}{D}
\ncline{-}{C}{E}
\ncline{-}{D}{E}
\end{pspicture}
}
\psscalebox{0.5}{
\begin{pspicture}(0,-0.5)(2.5,2)
{
\cnodeput(0,0){A}{}
\cnodeput(0,1.5){B}{}
\cnodeput(1.5,0){C}{}
\cnodeput(1.5,1.5){D}{}
\cnodeput(0.75,0.75){E}{}
}
\ncline{-}{A}{B}
\ncline{-}{B}{D}
\ncline{-}{D}{C}
\ncline{-}{C}{A}
\ncline{-}{A}{E}
\ncline{-}{B}{E}
\ncline{-}{C}{E}
\ncline{-}{D}{E}
\end{pspicture}
}}
\subfigure[m=9]{
\psscalebox{0.5}{
\begin{pspicture}(-1,-1)(3,2)
{
\cnodeput(0,0){A}{}
\cnodeput(0,1){B}{}
\cnodeput(1,0){C}{}
\cnodeput(1,1){D}{}
\cnodeput(2,0.5){E}{}
}
\ncline{-}{A}{B}
\ncline{-}{A}{C}
\ncline{-}{A}{D}
\ncline{-}{B}{C}
\ncline{-}{B}{D}
\ncline{-}{C}{D}
\ncline{-}{B}{E}
\ncline{-}{C}{E}
\ncline{-}{D}{E}
\end{pspicture}
}}
\subfigure[m=10]{
\psscalebox{0.5}{
\begin{pspicture}(-1.5,-1.5)(1.5,2)
{
\cnodeput(.37,1.141){A}{}
\cnodeput(-0.97,0.705){B}{}
\cnodeput(-0.97,-0.705){C}{}
\cnodeput(0.37,-1.14){D}{}
\cnodeput(1.2,0){E}{}
}
\ncline{-}{A}{B}
\ncline{-}{A}{C}
\ncline{-}{A}{D}
\ncline{-}{A}{E}
\ncline{-}{B}{C}
\ncline{-}{B}{D}
\ncline{-}{B}{E}
\ncline{-}{C}{D}
\ncline{-}{C}{E}
\ncline{-}{D}{E}
\end{pspicture}
}
}
\caption{All connected networks with $5$ vertices, $4\leq m \leq 10$ edges and the largest diameter. The first graph of each group is also a maximum average distance graph.}
\label{MaxDiameter5VerticesAll}
\end{figure}
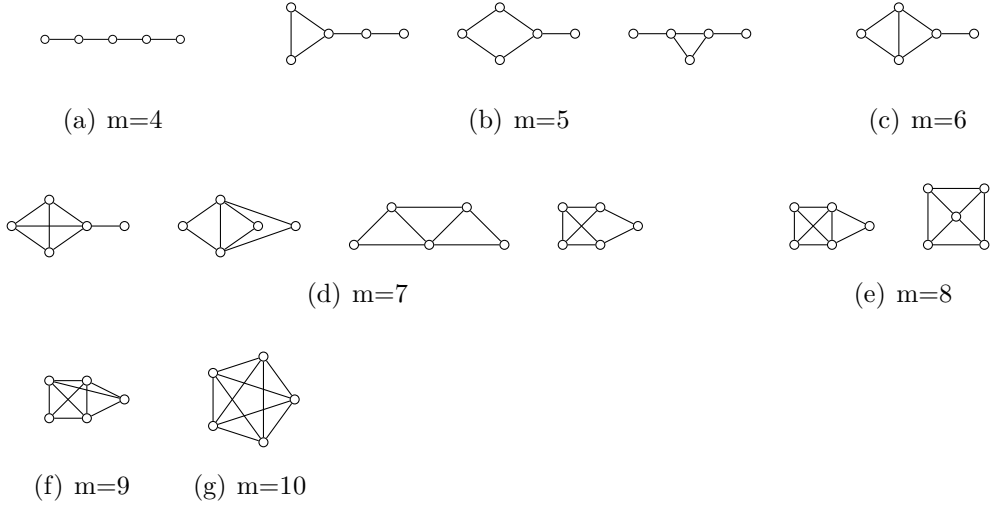

From the previous analysis we can conclude that the only tree with the largest diameter is the path graph.
A graph with the largest diameter is not necessarily unique for any size $N \leq m\leq {N \choose 2}-2$.
The list of all the graphs of order $N=5$ and size $4\leq m \leq 10$ are shown in Figure \ref{MaxDiameter5VerticesAll}.

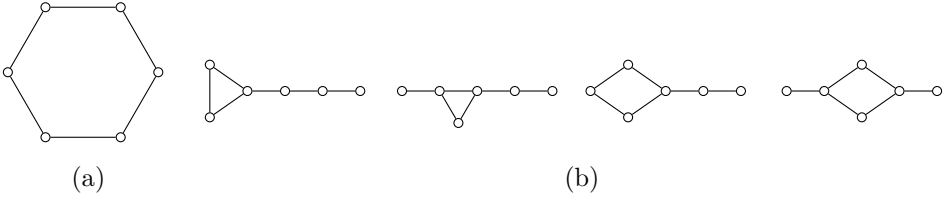
\begin{figure}[htb]
\subfigure[]{
\psscalebox{0.5}{
\begin{pspicture}(-1.8,-2)(2.2,2)
{
\cnodeput(1,1.73){A}{}
\cnodeput(-1,1.73){B}{}
\cnodeput(-2,0){C}{}
\cnodeput(-1,-1.73){D}{}
\cnodeput(1,-1.73){E}{}
\cnodeput(2,0){F}{}
}
\ncline{-}{A}{B}
\ncline{-}{B}{C}
\ncline{-}{C}{D}
\ncline{-}{D}{E}
\ncline{-}{E}{F}
\ncline{-}{F}{A}
\end{pspicture}
}
}
\subfigure[]{
\psscalebox{0.5}{
\begin{pspicture}(-0.2,-1.5)(4.3,1)
{
\cnodeput(0,-0.7){A}{}
\cnodeput(0,0.7){B}{}
\cnodeput(1,0){C}{}
\cnodeput(2,0){D}{}
\cnodeput(3,0){E}{}
\cnodeput(4,0){F}{}
}
\ncline{-}{A}{B}
\ncline{-}{A}{C}
\ncline{-}{B}{C}
\ncline{-}{C}{D}
\ncline{-}{D}{E}
\ncline{-}{E}{F}
\end{pspicture}
}
\psscalebox{0.5}{
\begin{pspicture}(-.4,-1.5)(4,1.5)
{
\cnodeput(0,0){A}{}
\cnodeput(1,0){B}{}
\cnodeput(1.5,-.85){C}{}
\cnodeput(2,0){D}{}
\cnodeput(3,0){E}{}
\cnodeput(4,0){F}{}
}
\ncline{-}{A}{B}
\ncline{-}{B}{C}
\ncline{-}{B}{D}
\ncline{-}{C}{D}
\ncline{-}{D}{E}
\ncline{-}{E}{F}
\end{pspicture}
}
\psscalebox{0.5}{
\begin{pspicture}(-.6,-1.5)(4.2,1)
{
\cnodeput(0,0){A}{}
\cnodeput(1,-0.7){B}{}
\cnodeput(1,0.7){C}{}
\cnodeput(2,0){D}{}
\cnodeput(3,0){E}{}
\cnodeput(4,0){F}{}
}
\ncline{-}{A}{B}
\ncline{-}{A}{C}
\ncline{-}{B}{D}
\ncline{-}{C}{D}
\ncline{-}{D}{E}
\ncline{-}{E}{F}
\end{pspicture}
}
\psscalebox{0.5}{
\begin{pspicture}(-1.6,-1.5)(3.8,1)
{
\cnodeput(0,0){A}{}
\cnodeput(1,-0.7){B}{}
\cnodeput(1,0.7){C}{}
\cnodeput(2,0){D}{}
\cnodeput(3,0){E}{}
\cnodeput(-1,0){F}{}
}
\ncline{-}{A}{B}
\ncline{-}{A}{C}
\ncline{-}{B}{D}
\ncline{-}{C}{D}
\ncline{-}{D}{E}
\ncline{-}{A}{F}
\end{pspicture}
}
}
\caption{\textbf{(a)} The only graph with $(N,m)=(6,6)$ and radius $R_{max}(6,6)=3$. \textbf{(b)} All graphs with the same order and size and maximum diameter, equal to $4$. No graph of the largest diameter coincides with the maximum radius network.}
\label{MaxRadiusVsMaxDiameter}
\end{figure}

Comparing the graphs with the largest radius with the graphs with the largest diameter, we find the following counterintuitive result:
\begin{corollary}
A network with the largest diameter does not necessarily have the largest radius.
Conversely, a network with the largest radius does not necessarily have the largest diameter.
\end{corollary}
An example is given in Figure \ref{MaxRadiusVsMaxDiameter}.

\FloatBarrier
\section{Resistance Distance}

The resistance distance between two vertices of a graph is equal to the total resistance between two  points of an electrical network, with each edge representing a unit resistance.

\begin{theorem}
The smallest possible resistance distance of a simple connected graph $\mathcal{G}=\mathcal{G}_{N,m}$ is
\begin{equation}
S_{min}(N,m)= \left\{ \begin{array}{ll}
\frac{2}{m-N+3} & N-1\leq m \leq 2N-3\\
\frac{2}{N}  & 2N-2 \leq m \leq {N \choose 2}.\\
\end{array} \right.
\end{equation}
\end{theorem}
\begin{proof}
If $m=N-1$, $\mathcal{G}$ is a tree, and it is not possible to have any resistors connected in parallel, so $\mathcal{S}_{min}(N,N-1)$=1.
Every tree is a minimum resistance distance graph as long as the endpoints of the circuit are two adjacent vertices.
When $N\leq m \leq 2N-3$, the network will have $c=m-N+1$ independent cycles.
In order to make the resistance as small as possible, we need to choose the endpoints of the circuit to be two adjacent nodes, and in addition to be parts of cycles as short as possible.
This is because the smaller the resistance of each branch of the cycle, the smaller the resistance between the two endpoints.
Since a cycle has at least $3$ vertices (two vertices cannot be connected with more than one edge, since the graphs are assumed to be simple), the cycles need to be of length $3$, and all of them have a common edge, the endpoints of which are the endpoints of the circuit.
As a result, the total resistance will be the combination of a unit resistor in parallel to $m-N+1$ pairs of resistors connected in parallel, which means that
\begin{equation}
S_{min}(N,m)=\frac{1}{1+\frac{1}{2}(m-N+1)}=\frac{2}{m-N+3}  \textrm{ ,}\quad N-1\leq m \leq 2N-3.
\end{equation}
Adding more edges to the network has no effect on the total impedance, since all vertices except for the circuit endpoints will have the same potential, equal to half of the voltage difference applied to the circuit ends (assuming that we have arbitrarily set the potential of one of the endpoints equal to zero).
Consequently, no current would flow among them, and we can write
\begin{equation}
S_{min}(N,m)=\frac{1}{1+\frac{1}{2}((2N-3)-N+1)}=\frac{2}{N}  \textrm{ ,}\quad  m\geq 2N-3.
\end{equation}
\end{proof}

\begin{corollary}
A graph with size $N-1 \leq m \leq 2N-3$ has minimum resistance distance if and only if it has a subgraph consisting of $m-N+1$ triangles that have one edge in common. The endpoints of the common edge are the endpoints of the circuit.
For $2N-2 \leq m \leq {N \choose 2}$, any graph with at least two vertices of full degree is a minimum resistance distance graph.
\end{corollary}

We now turn our attention to the structure of the networks with the largest resistance distance. We will start from networks of relatively large size (almost complete graphs) and then move on to finding their form in the general case.

\begin{lemma}
The maximum resistance of a network of fixed order is a decreasing function of its size.
\begin{equation}
S_{max}(N,m) \geq S_{max}(N,m+g) \quad \textrm{for } g \geq 0
\end{equation}
\label{MaxResistanceDecreasingFunction}
\end{lemma}
\begin{proof}
If $S_{max}(N,m)<S_{max}(N,m+g)$, we remove $g$ edges from the network of larger size.
This cannot decrease the total resistance, since there is now no voltage drop between the vertices that were previously connected.
Therefore, the resulting graph will have total resistance larger than the resistance of the initial network of smaller size, which contradicts the hypothesis.
\end{proof}

\begin{figure}[htb]
\subfigure[]{
\psscalebox{0.4}{
\begin{pspicture}(-4.5,-6)(4.5,5)
{
\cnodeput[fillstyle=solid,fillcolor=blue](3.83,3.21){A}{\strut}
\cnodeput[fillstyle=solid,fillcolor=green](0.87,4.92){B}{\strut}
\cnodeput[fillstyle=solid,fillcolor=blue](-2.5,4.33){C}{\strut}
\cnodeput[fillstyle=solid,fillcolor=yellow](-4.7,1.71){D}{\strut}
\cnodeput[fillstyle=solid,fillcolor=yellow](-4.7,-1.71){E}{\strut}
\cnodeput[fillstyle=solid,fillcolor=blue](-2.5,-4.33){F}{\strut}
\cnodeput[fillstyle=solid,fillcolor=red](0.87,-4.92){G}{\strut}
\cnodeput[fillstyle=solid,fillcolor=yellow](3.83,-3.21){H}{\strut}
\cnodeput[fillstyle=solid,fillcolor=yellow](5,0){I}{\strut}
}
\ncline{-}{A}{B}
\ncline{-}{A}{C}
\ncline{-}{A}{D}
\ncline{-}{A}{E}
\ncline{-}{A}{F}
\ncline{-}{A}{G}
\ncline{-}{A}{H}
\ncline{-}{A}{I}
\ncline{-}{C}{D}
\ncline{-}{C}{E}
\ncline{-}{C}{F}
\ncline{-}{C}{G}
\ncline{-}{C}{H}
\ncline{-}{C}{I}
\ncline{-}{D}{E}
\ncline{-}{D}{F}
\ncline{-}{D}{G}
\ncline{-}{D}{H}
\ncline{-}{D}{I}
\ncline{-}{E}{F}
\ncline{-}{E}{G}
\ncline{-}{E}{H}
\ncline{-}{E}{I}
\ncline{-}{F}{G}
\ncline{-}{F}{H}
\ncline{-}{F}{I}
\ncline{-}{G}{H}
\ncline{-}{G}{I}
\ncline{-}{H}{I}
\ncline{-}{B}{C}
\ncline{-}{B}{F}
\end{pspicture}
}
}
\subfigure[]{
\psscalebox{0.4}{
\begin{pspicture}(-8,-1)(8.5,12)
{
\cnodeput[fillstyle=solid,fillcolor=blue](-6,8){A}{\strut}
\cnodeput[fillstyle=solid,fillcolor=blue](0,8){B}{\strut}
\cnodeput[fillstyle=solid,fillcolor=yellow](-4.5,4){C}{\strut}
\cnodeput[fillstyle=solid,fillcolor=yellow](-1.5,4){D}{\strut}
\cnodeput[fillstyle=solid,fillcolor=yellow](1.5,4){E}{\strut}
\cnodeput[fillstyle=solid,fillcolor=yellow](4.5,4){F}{\strut}
\cnodeput[fillstyle=solid,fillcolor=red](0,0){G}{\strut}
\cnodeput[fillstyle=solid,fillcolor=blue](6,8){H}{\strut}
\cnodeput[fillstyle=solid,fillcolor=green](0,12){I}{\strut}
}
\ncline{-}{A}{C}
\ncline{-}{A}{D}
\ncline{-}{A}{E}
\ncline{-}{A}{F}
\ncline{-}{A}{G}
\ncline{-}{B}{C}
\ncline{-}{B}{D}
\ncline{-}{B}{E}
\ncline{-}{B}{F}
\ncline{-}{B}{G}
\ncline{-}{C}{G}
\ncline{-}{D}{G}
\ncline{-}{E}{G}
\ncline{-}{F}{G}
\ncline{-}{H}{C}
\ncline{-}{H}{D}
\ncline{-}{H}{E}
\ncline{-}{H}{F}
\ncline{-}{H}{G}
\ncline{-}{I}{A}
\ncline{-}{I}{B}
\ncline{-}{I}{H}
\end{pspicture}
}
}
\subfigure[]{
\psscalebox{0.4}{
\begin{pspicture}(-2,-1)(2.5,12)
{
\cnodeput[fillstyle=solid,fillcolor=green](0,12){A}{\strut}
\cnodeput[fillstyle=solid,fillcolor=blue](0,8){B}{\strut}
\cnodeput[fillstyle=solid,fillcolor=yellow](0,4){C}{\strut}
\cnodeput[fillstyle=solid,fillcolor=red](0,0){D}{\strut}
}
\ncline{-}{A}{B}
\ncline{-}{B}{C}
\ncline{-}{C}{D}
\ncarc[linewidth=1pt,arcangle=65]{B}{D}
\rput(-1,10){\rnode{R0}{\Huge{$\frac{1}{\alpha}$}}}
\rput(-1.9,6){\rnode{R1}{\Huge{$\frac{1}{\alpha(C-1-\alpha)}$}}}
\rput(-1.5,2){\rnode{R2}{\Huge{$\frac{1}{C-1-\alpha}$}}}
\rput(2.7,4){\rnode{R3}{\Huge{$\frac{1}{a}$}}}
\end{pspicture}
}
}
\caption{\textbf{(a)} An almost complete graph has the largest resistance distance among all other graphs of the same order and size. The endpoints with that resistance are the peripheral vertex (green) and an arbitrary central vertex that is \textit{not} connected to it (red). The central vertices that the peripheral vertex connects to (blue), all have the same potential. All central vertices that are not connected to the peripheral vertex (yellow) also have the same potential. \textbf{(b)} Taking into consideration the symmetry of the circuit, we can remove the edges among vertices that belong to the same group. \textbf{(c)} The previous circuit can be simplified by collapsing all nodes of one group to one ``super-node'', and considering all resistors that connect different groups to be connected in parallel. The weights of the edges correspond to the resistances among the ``super-nodes''.}
\label{AlmostCompleteResistanceTransformation}
\end{figure}
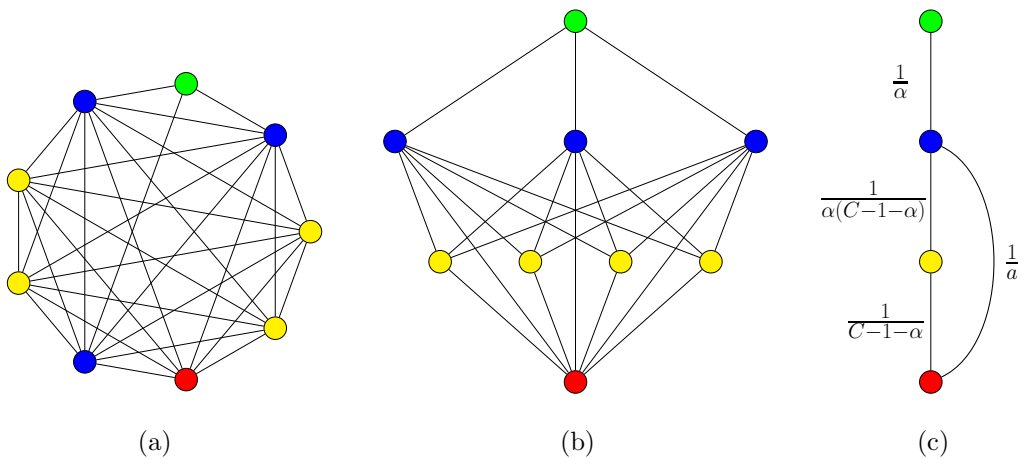

\begin{lemma}
The largest possible resistance distance of a network with $m\geq {N-1 \choose 2}+1$ is
\begin{equation}
S_{max}(N,m)=\frac{\alpha+C +1}{\alpha C}
\end{equation}
where
\begin{equation}
C=N-1 \quad \textrm{and} \quad \alpha=m-{N-1 \choose 2}.
\end{equation}
\label{AlmostCompleteGraphsMaxResistance}
\end{lemma}
\begin{proof}
The almost complete graph will have the largest possible resistance when the endpoints of the circuit are the peripheral vertex and a central vertex it is \textit{not} connected to.
The reason is that any other graph will have more shorter paths to the ground vertex, and thus smaller total resistance \cite{RandomWalkElectricalNetworks}.
In the almost complete graph, because of the symmetry, all the vertices that are neighbors of the peripheral vertex have the same potential.
Similarly, all vertices that are \textit{not} neighbors of the peripheral vertex, except for the ground vertex also have the same potential.
The ground vertex has zero potential.
Thus, we may remove the edges between the $\alpha$ neighbors of the peripheral vertex, and the $N-2-\alpha$ non-neighbors of the peripheral vertex, excluding the ground vertex.
Then, we merge the edges that connect these three sets of vertices, as if they were connected in parallel.
The result is shown in Figure \ref{AlmostCompleteResistanceTransformation}.
The total resistance of the transformed network is
\begin{equation}
\begin{aligned}
S_{max}(N,m) &=\frac{1}{\alpha}+\frac{1}{\frac{1}{\frac{1}{\alpha}}+\frac{1}{\frac{1}{\alpha(N-2-\alpha)}+\frac{1}{N-2-\alpha}}} \\
 &=\frac{1}{\alpha}+\frac{1}{\alpha +\frac{\alpha(N-2-\alpha)}{\alpha+1}} \\
 &=\frac{N+\alpha}{\alpha(N-1)}. \\
\end{aligned}
\end{equation}
\end{proof}

\begin{lemma}
The maximum resistance of a network $\mathcal{G}(N,m)$ can be at most one larger than the maximum resistance $S_{max}(N-1,m-1)$, provided that both networks are connected. More precisely
\begin{equation}
S_{max}(N,m) \leq S_{max}(N-1,m-1)+1.
\end{equation}
\label{GraphResistanceIncrementalInequality}
\end{lemma}
\begin{proof}
We start from the network $\mathcal{H}(N-1,m-1)$ with the largest resistance, and add one extra vertex with unit degree to one of its endpoint vertices.
If the Lemma does not hold, then the resulting graph $\mathcal{G}=\mathcal{G}(N,m)$ will have resistance 
\begin{equation}
S(\mathcal{G})=1+S_{max}(N-1,m-1)<S_{max}(N,m).
\label{MaxResistanceContradictingAssumption}
\end{equation}
Now assume that the network $\mathcal{K}$ with the maximum resistance has vertex $u$ with degree $d_{u}$ as an endpoint.
If $d_{u}=1$, then network $\mathcal{H}$ was not a maximum resistance network and the Lemma is proved.
If $d_{u}>1$, then the potential difference between $u$ and all its neighbors will be smaller or equal to one, since the current flowing through $u$ will be divided among all the resistors that are adjacent to $u$.
Consequently, removing vertex $u$ along with its edges, we can find a relation with the resulting network $\mathcal{M}$
\begin{equation}
\begin{aligned}
S_{max}(N,m) &\leq \epsilon + S(\mathcal{M})
&\leq \epsilon + S_{max}(N-1,m-d_{u})
\end{aligned}
\label{ResistanceAfterRemovingAGeneralVertex}
\end{equation}
with $0 \leq \epsilon \leq 1$.
Combining equations \eqref{MaxResistanceContradictingAssumption}  and \eqref{ResistanceAfterRemovingAGeneralVertex}, it is evident that 
\begin{equation}
S_{max}(N-1,m-1) <S_{max}(N-1,m-d_{u})-(1-\epsilon)
\end{equation}
which, according to Lemma \ref{MaxResistanceDecreasingFunction}, is not possible.
\end{proof}

\begin{theorem}
The largest possible resistance distance of a simple connected graph $\mathcal{G}=\mathcal{G}_{N,m}$ is
\begin{equation}
S_{max}(N,m)=P-1+\frac{\alpha+C +1}{\alpha C}
\end{equation}
where $C,P$ and $\alpha$ are defined in equations \eqref{MaxDistanceGraphCliqueOrder}, \eqref{MaxDistanceGraphPathOrder} and \eqref{CutEdgesBetweenPathAndClique} respectively.
\end{theorem}
\begin{proof}
Using the Lemmas above, we can analytically compute the largest resistance, and find the form of the graphs that have it.
We repeatedly apply Lemma \ref{GraphResistanceIncrementalInequality}, until we are left with an almost complete graph, at which point we can make use of Lemma \ref{AlmostCompleteGraphsMaxResistance}.
\begin{equation}
S_{max}(N,m)\leq k+ S_{max}(N-k,m-k).
\end{equation}
We can achieve the equality in the last equation by connecting a new vertex with unit degree to one of the endpoints of the previous graph, such that
\begin{equation}
k=P-1 \quad \textrm{and} \quad N-k=C+1
\end{equation}
where the total resistance is the sum of a path graph with $P$ vertices, and an almost complete graph with the respective values of $C$ and $\alpha$ (see also condition \eqref{DiameterStoppingCondition}).
The form of these maximum resistance networks are easy to identify: We recursively connect in series a unit resistor in one of the two endpoints of the previous circuit.
\end{proof}

\begin{figure}[htb]
\includegraphics[scale=0.37]{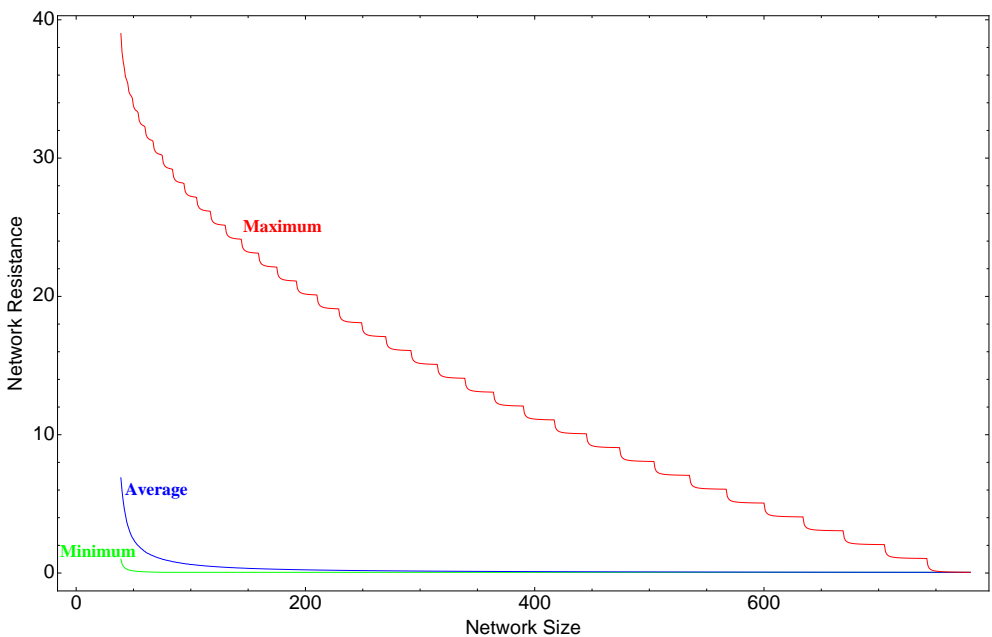}
\caption{The largest, smallest and average resistance of graphs of order $N=40$ and size $39\leq m \leq 780$. The minimum and maximum bounds are analytically computed. The average resistance of each graph is approximated by the mean of $10^{4}$ random graphs of the respective size. The statistical average of a network of relatively small size is very close to the minimum resistance.}
\label{MinAvgResistanceExample}
\end{figure}

An example showing the minimum, average and maximum resistance for networks of order $N=40$ and increasing size is shown in Figure \ref{MinAvgResistanceExample}.

\begin{corollary}
The number of non-isomorphic graphs with the largest resistance distance is
\begin{equation}
I_{S_{max}}(N,m)= \left\{ \begin{array}{ll}
P+1 & \alpha=1 \textrm{ or } \alpha =C-1 \\
\Big \lceil \frac{P+1}{2}  \Big \rceil & 2\leq \alpha \leq C-2 \\
\end{array} \right.
\end{equation}
where $C, P$ and $\alpha$ are defined in equations \eqref{MaxDistanceGraphCliqueOrder}, \eqref{MaxDistanceGraphPathOrder} and \eqref{CutEdgesBetweenPathAndClique} respectively.
\end{corollary}
\begin{proof}
The order in which we place linear resistors in series has no effect on the total resistance of the circuit.
We can serially place $n_{1}$ and $n_{2}$ resistors at each side of the almost complete subgraph, such that
\begin{equation}
n_{1}+n_{2}=P \quad \textrm{with } n_{1}\geq 0 \textrm{ and } n_{2} \geq 0.
\end{equation}
The last equation has $P+1$ solutions.
When $a=1$ or $a=C-1$, then the almost complete graph is symmetric, which means that we count every non-isomorphic graph twice.
Adjusting for this special case the number of non-isomorphic graphs, we get the desired result.
\end{proof}

\begin{corollary}
The networks with the largest average distance have the largest resistance distance. The converse does not always hold.
\end{corollary}

\FloatBarrier

\section{Sensitivity to Rewiring}

In most cases when designing a network, there are many constraints that need to be satisfied, and the properties discussed so far are only proxies to determining other desirable qualities of a network.
Also it is reasonable to expect that our networks may not be allowed to have the extremal values of the properties described in the previous sections, especially given that sometimes there are conflicting requirements for the network function.
Under these considerations, we are interested in knowing how robust these structures are, in other words, how sensitive these properties are to changes in the interconnection patterns.

\begin{figure}[tb]
\subfigure[]{
\includegraphics[scale=0.2]{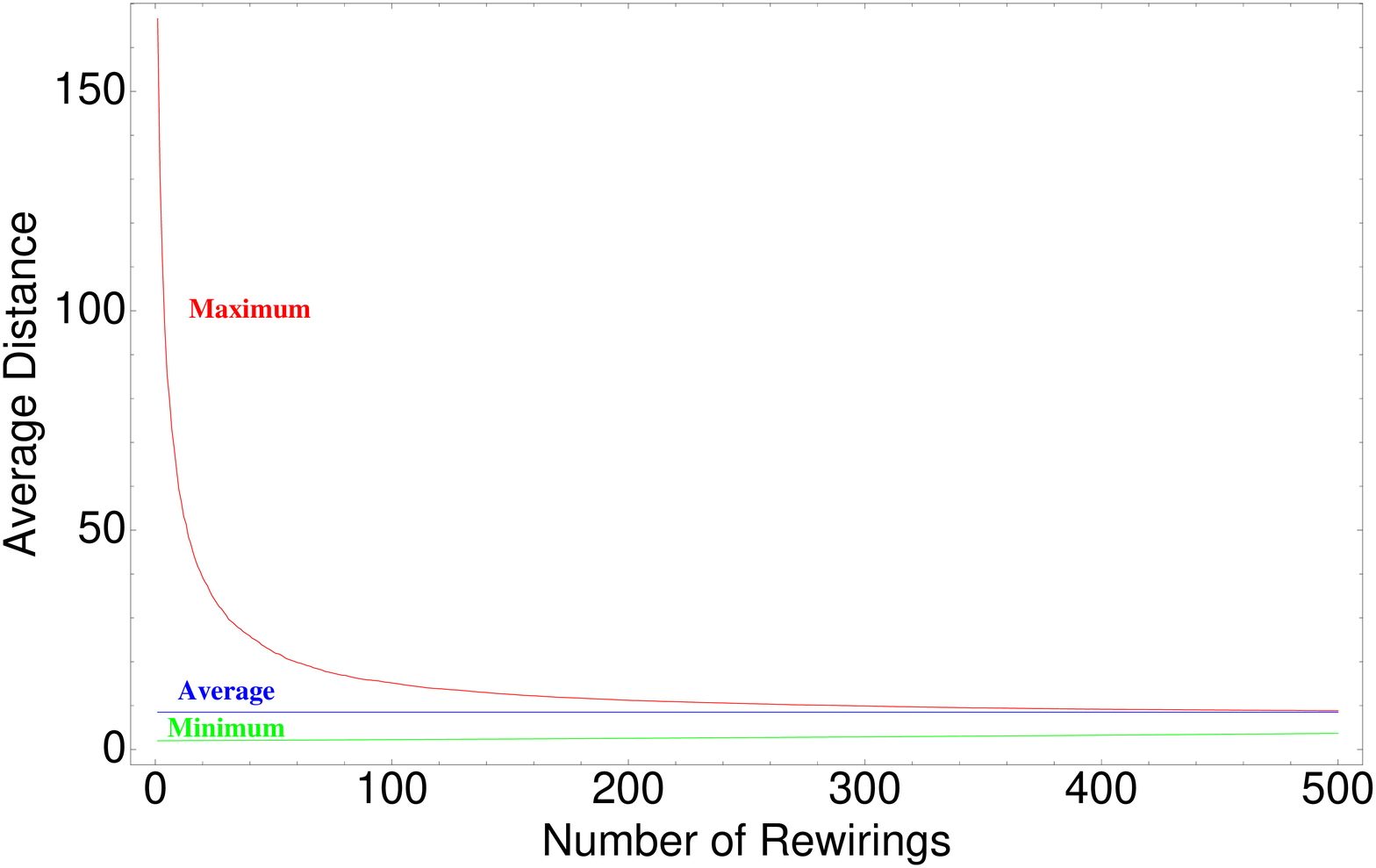}
}
\subfigure[]{
\includegraphics[scale=0.2]{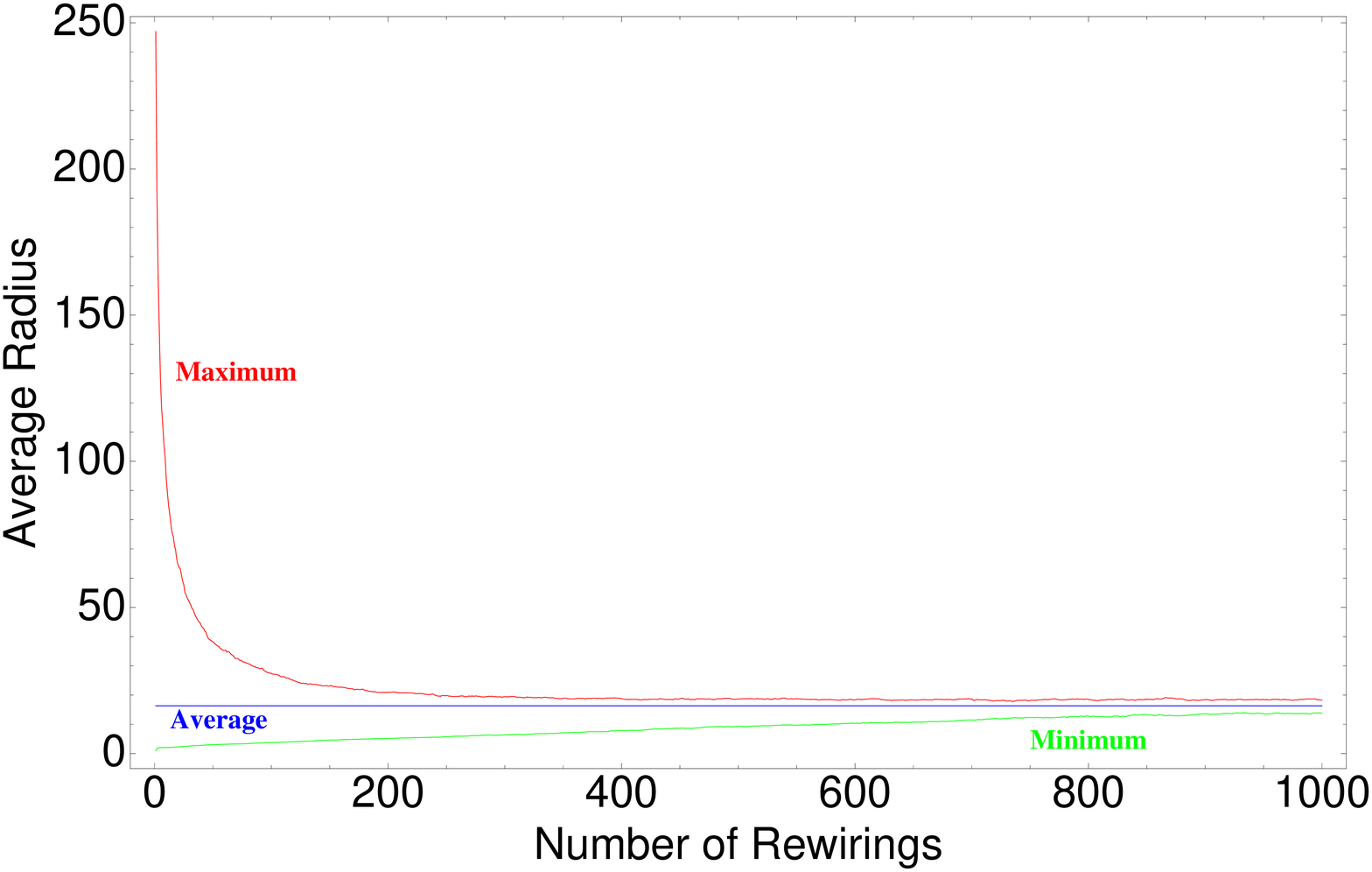}
}
\subfigure[]{
\includegraphics[scale=0.2]{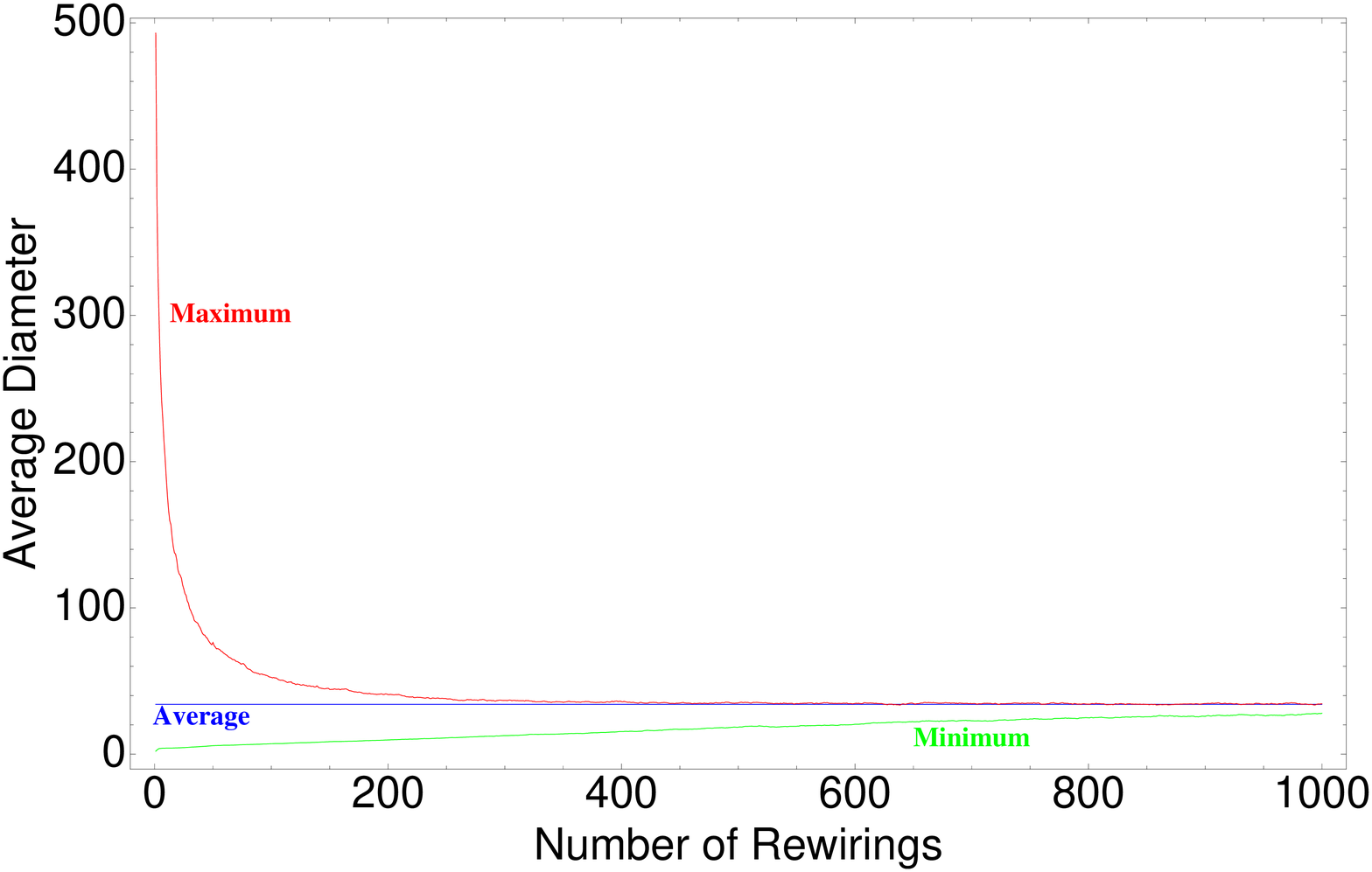}
}
\subfigure[]{
\includegraphics[scale=0.2]{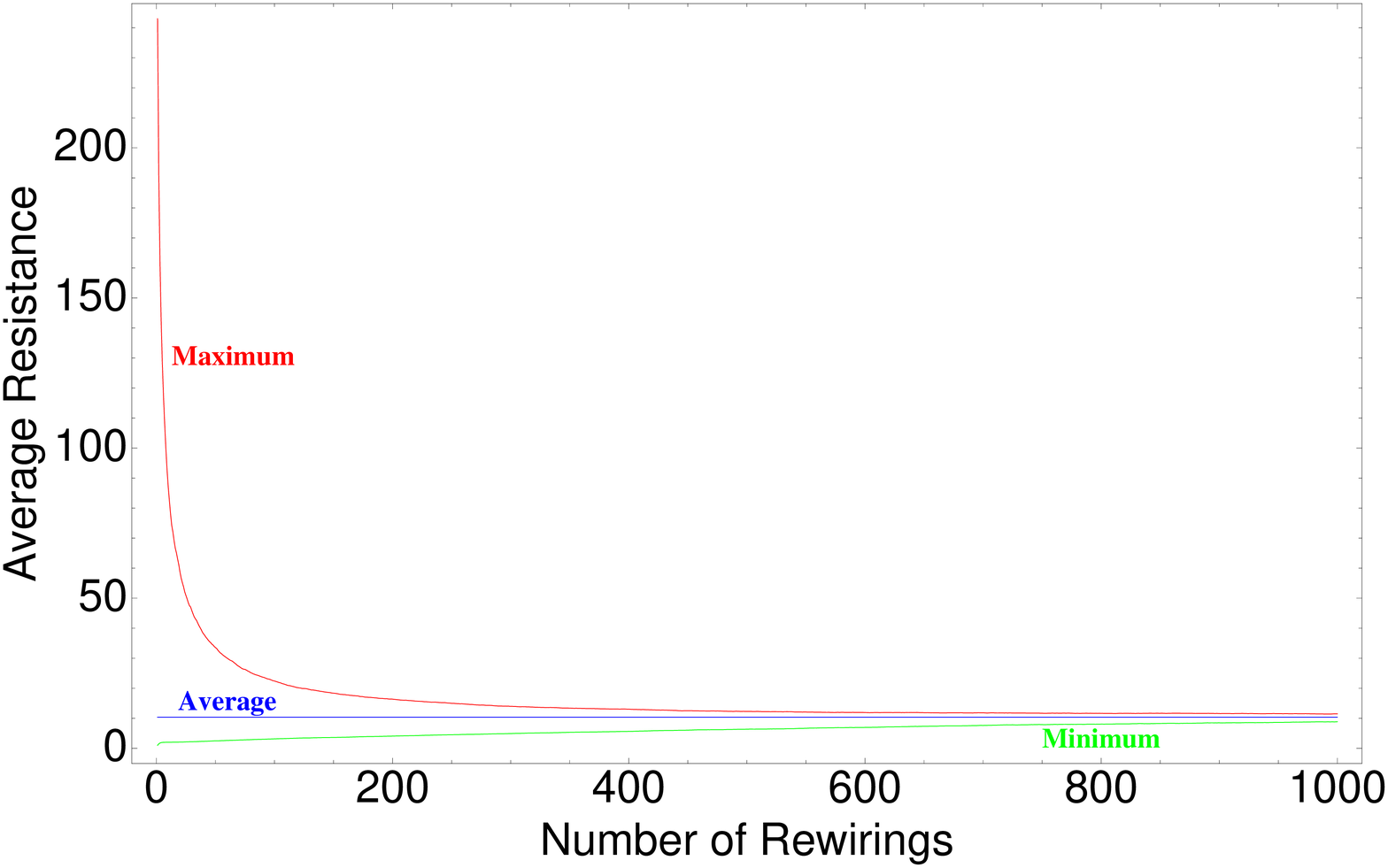}
}
\caption{Evolution of the average distance, radius, diameter and resistance of networks after successive random rewirings. Each curve is the average of 100 experiments, rewiring networks of the same size and order $N=500$. The curves for the average shortest path length correspond to a network of size $m=600$. The curves for the evolution of radius, diameter and resistance correspond to networks of size $m=525$.}
\label{StatisticalAnalysis}
\end{figure}

To this end, we can start with two networks with the extremal properties, and keep rewiring one of the edges, making sure that the network remains connected.
At every step, we measure the properties of each network instance, and monitor its evolution as we introduce more and more randomness in their architecture.
Eventually, both will resemble a random graph, with the average distance, betweenness centrality, efficiency, radius,  diameter and resistance being close to the statistical average.
The question though is how fast this state is reached.
If the change is very fast, this means that a few changes in the structure are able to negate the advantages that an extremal graph is able to provide.
However, if a particular property does not change much after several rewirings, we can afford to build networks with many other desirable properties without the need to follow exactly the specific architecture for every given property.

We have tested the change of average distance, radius, diameter and resistance for networks of small order.
Refering to Figure \ref{StatisticalAnalysis}, we find that the average distance of networks with large average shortest path length decreases very quickly after a few rewirings.
Convergence to the statistical average is much slower for networks that start with the smallest average distance. 
The number of rewirings required to get close to the statistical average is an increasing function of the order of the network when the edge density is constant, as one would expect, since there are more edges that need to be rewired in order to make a difference.
The same general conclusions can be drawn for the evolution of the network radius, diameter and resistance.

The general conclusion is that networks with the smallest average distance, radius and diameter and resistance are much more robust to changes in their architecture because they rely on a global pattern of interconnections, and each edge has a small role in ensuring that property, so its  conservation is diffused among many edges.
On the contrary, networks  with maximum distance, radius, diameter and resistance are very sensitive to change, because most vertex connections are prohibited, in the sense that if currently non-neighboring vertices are connected, there is a dramatic change towards the statistical average.

\FloatBarrier

\end{document}